\documentclass[superscriptaddress,amsmath,amssymb,aps,prb,floatfix,longbibliography]{revtex4-1}
\usepackage[utf8]{inputenc}
\usepackage{graphicx}
\usepackage{dcolumn}
\usepackage{bm}
\usepackage{hyperref}
\usepackage{amsmath,amsfonts,amssymb, mathtools} 
\usepackage[dvipsnames]{xcolor}
\usepackage{pdfpages}
\usepackage{float}
\usepackage{braket}
\usepackage{tikz}
\usepackage{CJK}
\usepackage{siunitx}
\usepackage{tabularx}
\usepackage[normalem]{ulem}
\makeatletter
\AtBeginDocument{\let\LS@rot\@undefined}
\makeatother

\newcommand{\Kt}{\ensuremath \kappa_\mathrm{t}} \newcommand{\Kb}{\ensuremath \kappa_\mathrm{b}}
\newcommand{\Ktprime}{\ensuremath \kappa'_\mathrm{t}}
\newcommand{\Kbprime}{\ensuremath \kappa'_\mathrm{b}}
 
 \newcommand{\kFvec}{\vec{k}_\mathrm{F}}
\newcommand{\Q}{\vec{Q}}
\newcommand{\twist}{\ensuremath \theta=2.37^\circ} 
\newcommand{\twisttheory}{\ensuremath \theta=2.3^\circ}

\newcommand{\me}{\ensuremath m_\mathrm{e}} 
\newcommand{\Rxx}{\ensuremath R_\mathrm{xx}}  

\newcommand{\Dmax}{D=\SI{0.47}{V/nm}} 
\newcommand{\Vsd}{\ensuremath V_\mathrm{sd}} 
\newcommand{\Vt}{\ensuremath V_\mathrm{t}}    \newcommand{\Vb}{\ensuremath V_\mathrm{b}} 
\newcommand{\Ct}{\ensuremath C_\mathrm{t}}    \newcommand{\Cb}{\ensuremath C_\mathrm{b}} \newcommand{\Cm}{\ensuremath C_\mathrm{BLG}}

\newcommand{\ns}{\ensuremath n_\mathrm{s}}   
  
\newcommand{\Bperp}{\ensuremath B_\mathrm{\perp}}   
\newcommand{\Bparallel}{\ensuremath B_\mathrm{\parallel}}  \newcommand{\Bc}{\ensuremath B_\mathrm{c}}    
\newcommand{\muB}{\ensuremath \mu_\mathrm{B}} 

\newcommand{\Ez}{\ensuremath E_\mathrm{z}}

\newcommand{\gv}{\ensuremath g_\mathrm{v}}
\newcommand{\Deltaup}{\ensuremath \Delta_\mathrm{\uparrow}}
\newcommand{\Deltadown}{\ensuremath \Delta_\mathrm{\downarrow}}
\newcommand{\DeltaV}{\ensuremath \Delta_\mathrm{V}}

\newcommand{\Dt}{\ensuremath D_\mathrm{t}}  \newcommand{\Db}{\ensuremath D_\mathrm{b}}
\newcommand{\Dblg}{\ensuremath D_\mathrm{BLG}} 
\newcommand{\Deltat}{\ensuremath \Delta_\mathrm{t}}
\newcommand{\Deltab}{\ensuremath \Delta_\mathrm{b}}
\newcommand{\DeltaCF}{\ensuremath \Delta_\mathrm{0}}
\newcommand{\Deltazero}{\ensuremath \Delta_\mathrm{0}}
\newcommand{\Deltabands}{\ensuremath \Delta_\mathrm{v}}
\newcommand{\ndouble}{\ensuremath n_\mathrm{double}}
\newcommand{\nsingle}{\ensuremath n_\mathrm{single}}
\newcommand{\Edouble}{\ensuremath E_\mathrm{double}}
\newcommand{\Esingle}{\ensuremath E_\mathrm{single}}
\newcommand{\Cq}{\ensuremath C_\mathrm{q}}
\newcommand{\eps}{\ensuremath \epsilon_0}

\newcommand{\meff}{\ensuremath m^*}

\newcommand{\probk}{\mathrm{Prob}(\Kt)}

\begin{document}
	\title{Correlated electron-hole State in Twisted Double Bilayer Graphene}
	
	\author{Peter Rickhaus}
	\email{peterri@phys.ethz.ch}
	\author{Folkert K. de Vries}
	\affiliation{Solid State Physics Laboratory, ETH Zürich,~CH-8093~Zürich, Switzerland}
	\author{Jihang Zhu}
	\affiliation{Department of Physics, University of Texas at Austin, Austin, Texas 78712, USA}
	\author{El\'ias Portolés}	
	\author{Giulia Zheng}
	\author{Michele Masseroni}
	\author{Annika Kurzmann}
	\affiliation{Solid State Physics Laboratory, ETH Zürich,~CH-8093~Zürich, Switzerland}
	\author{Takashi Taniguchi}
	\author{Kenji Wantanabe}
	\affiliation{National Institute for Material Science, 1-1 Namiki, Tsukuba 305-0044, Japan}
	\author{Allan H. MacDonald}
	\affiliation{Department of Physics, University of Texas at Austin, Austin, Texas 78712, USA}
	\author{Thomas Ihn}
	\author{Klaus Ensslin}
	\affiliation{Solid State Physics Laboratory, ETH Zürich,~CH-8093~Zürich, Switzerland}
	
	\date{\today}
	
	
	\begin{abstract}
When twisted to angles near $1^\circ$, graphene multilayers provide a window on electron correlation physics. Here we report the discovery of a  correlated electron-hole state in double bilayer graphene twisted to $2.37^\circ$. At this angle the moiré states retain much of their isolated bilayer character, allowing their bilayer projections to be separately controlled by gates. We use this property to generate an energetic overlap between narrow isolated electron and hole bands with good nesting properties. Our measurements reveal the formation of ordered states with reconstructed Fermi surfaces, consistent with a density-wave state. This state can be tuned without introducing chemical dopants, enabling studies of correlated electron-hole states and their interplay with superconductivity.   
	\end{abstract}

	\maketitle

\begin{figure}
\centering
\includegraphics[width=0.5\textwidth]{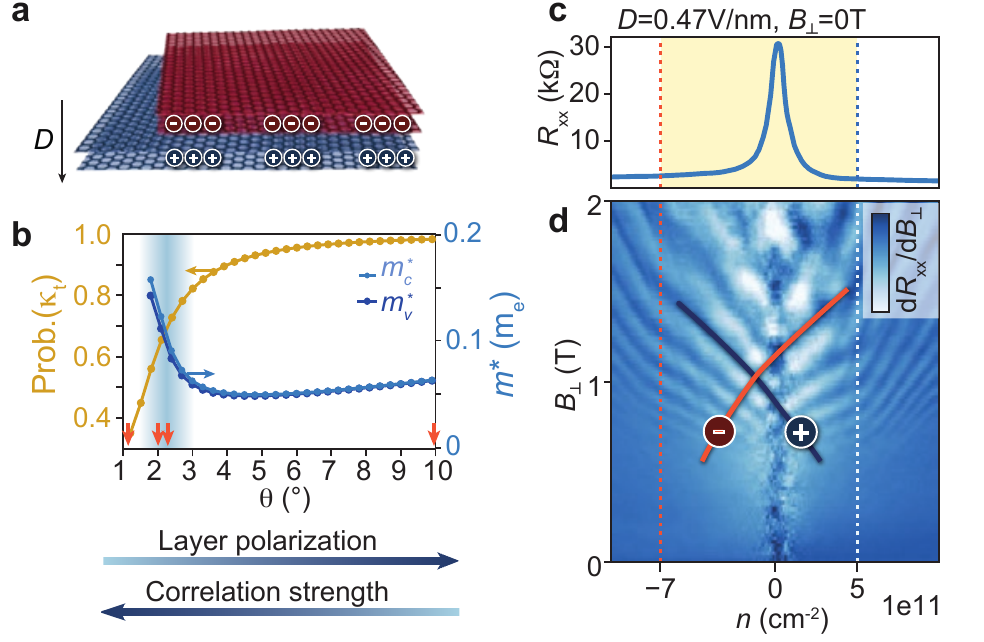}
    \caption{\textbf{Designed density-wave state in a moir\'e double bilayer.} 
	a) Applying an electric field $D$ to twisted double bilayer graphene will charge the top/bottom bilayer with electrons/holes. 
	b) The layer polarization is given by $\probk$ and the correlation strength scales with the effective mass $\meff$. $m^*_c$ ($m^*_v$) is the effective mass of conduction (valence) band. $\probk(\theta)$ and $\meff(\theta)$  (at layer energy difference $\DeltaV=\SI{30}{meV}$, see below) are obtained from band structure calculations at $\twisttheory$, showing that twists in the intermediate range (shaded) combine strong polarizations and large masses and are favorable for DW formation. The red arrows mark twist angles at which measurements in the SI, section \ref{sec:othertwist} and \ref{sec:twist2} are presented. A correlated electron-hole gap is seen  at intermediate twist angles of $\twist$ and $\theta=1.98^\circ$.
	c) A peak in $\Rxx$ occurs at $\Dmax$ and carrier density $n=0$, in a TDBG device with $\twist$, and is consistent 
	with a thermally activated gap (see SI section \ref{sec:thermal}).
	d) The bilayer's Landau-fan diagram demonstrates that the gap appears inside a regime with both electron and hole bands, as testified by the presence of Shubnikov-de Haas resistance peaks (orange and blue) that have opposite $n$ {\it vs.} $\Bperp$ slopes. Measurements are performed at $T=\SI{1.2}{K}$.
     }\label{fig:1}
\end{figure}

Fermi surface nesting refers to electron and hole Fermi surfaces that map onto each other under translation by a nesting wavevector $\Q$. Because nesting implies a small band energy cost for coherent superposition between electrons and holes, it favors interaction-driven broken symmetry states.
The nesting condition ($\epsilon(\kFvec+\Q)=-\epsilon(\kFvec))$, where epsilon is a small energy offset from the Fermi energy) implies \cite{Khomoskii2010} that if two closed Fermi surfaces are perfectly nested, they enclose the same area.
Mixing two Fermi surfaces nested by wavevector $\Q$ leads to density-wave (DW) order with wavelength $2\pi/Q$. In the  seminal theoretical work on DW states by Peierls \cite{Peierls1930}, nesting occurs between like-spins in half-filled bands, the interactions are lattice mediated, and the DW is accompanied by a lattice distortion.  
The Peierls transition is one of the first instabilities of the metallic state to be recognized, and has been observed in a large range of materials. 
DW states can also be favored by Coulomb interactions between electrons, in which case the lattice distortions\cite{Li2019} play a parasitic role only. 
The DW is then often referred to as an excitonic insulator\cite{Jerome1967}. This term is suggested by viewing the order as a condensation of bosonic electron-hole pair states.  Evidence for equilibrium excitonic condensation has been reported in 1T‐TiSe$_2$\cite{Kogar2017}, in Sb nanoflakes \cite{Li2019} and in double quantum wells at high magnetic fields \cite{Spielman2000,Du2017a}. Furthermore, significant progress towards quasi-equilibrium exciton condensation has been achieved in a MoSe$_2$/WSe$_2$ heterostructure under the application of a large interlayer bias.\cite{Wang2019} Condensation of non-equilibrium excitons and polaritons in optically pumped electron-hole fluids has also been studied bextensively.\cite{OpticalXC}
Correlated electron-hole states continue to attract attention due to their rich intrinsic physics and their close relationship to superconductivity \cite{Chang2012}.

Twisting Van der Waals (VdW) materials, including graphene bilayers (TBG) \cite{Cao2016,Cao2018a,Sharpe2019,Yankowitz2019,Stepanov2019,Saito2019} and double Bernal bilayers (TDBG) \cite{Koshino2019,Chebrolu2019, Choi2019, Liu2019,Shen2019, Burg2019,He2020}, 
is a proven strategy to engineer moir\'e bands that favor strongly correlated electronic states.  
In the present work we seek to realize electron-hole bands that are both nested and relatively narrow. To this end we have studied the properties of TDBG \cite{Koshino2019,Chebrolu2019, Choi2019, Liu2019,Shen2019, Burg2019,He2020}  at intermediate twist angles, where the layer coupling is strong enough to form moir\'e bands, but weak enough to retain the polarizability of decoupled bilayers \cite{Lucian2011,Sanchez2012,TutucXC,DeVries2020}. 
This polarizability can be quantified by the probability $\probk$ to find a band state at the $\Kt$-point (Brillouin zone corner of the top bilayer) in the top bilayer. If $\probk$ is large, which is the case at large twist angles, applying a displacement field $D$ will charge the top bilayer with electrons and the bottom bilayer with holes (Fig.~\ref{fig:1}a). But correlated electron-hole states are fragile and if they occur, only indirectly observable \cite{TutucXC}. Thus the correlation strength needs to be maximized by approaching smaller angles where the narrowed moir\'e bands have a larger effective mass $\meff$ (Fig.~\ref{fig:1}b). In SI \ref{sec:theorymeff}, we discuss the $\probk$ and $\meff$ dependence on twist angle and external field using the Bistritzer-MacDonald model\cite{MATBG} which is extended to the TDBG case (SI \ref{sec:model_H}).

Here we show that correlated electron-hole states are favored to form in the intermediate twist angle regime, roughly between $2^\circ$ and $3^\circ$. Figs.~\ref{fig:1}cd summarize the main experimental findings at $\twist$. We observe a resistance peak at zero total density $n=0$ and displacement field $\Dmax$ (Fig.~\ref{fig:1}c) which appears when electrons and holes with approximately the same density coexist. This coexistence is evident in Landau-fan measurements (Fig.~\ref{fig:1}d) that reveal electron and hole minibands which cross in energy. In this report, we first discuss how these overlapping electron-hole regions are formed by tuning $D$ and $n$ and then show how the emerging correlated state is influenced by these parameters in the experiment. In a next step, we discuss the data in a parallel magnetic field and show that it is in agreement with Fermi surface nesting. The lifting of spin and valley degeneracy in parallel and perpendicular magnetic field allows us to address the spin or valley symmetry of the correlated electron-hole state. Finally,we support our interpretation by Hartree-Fock (HF) simulations and suggest that the correlated state can be viewed as a DW.

\begin{figure}
\centering
\includegraphics[width=1\textwidth]{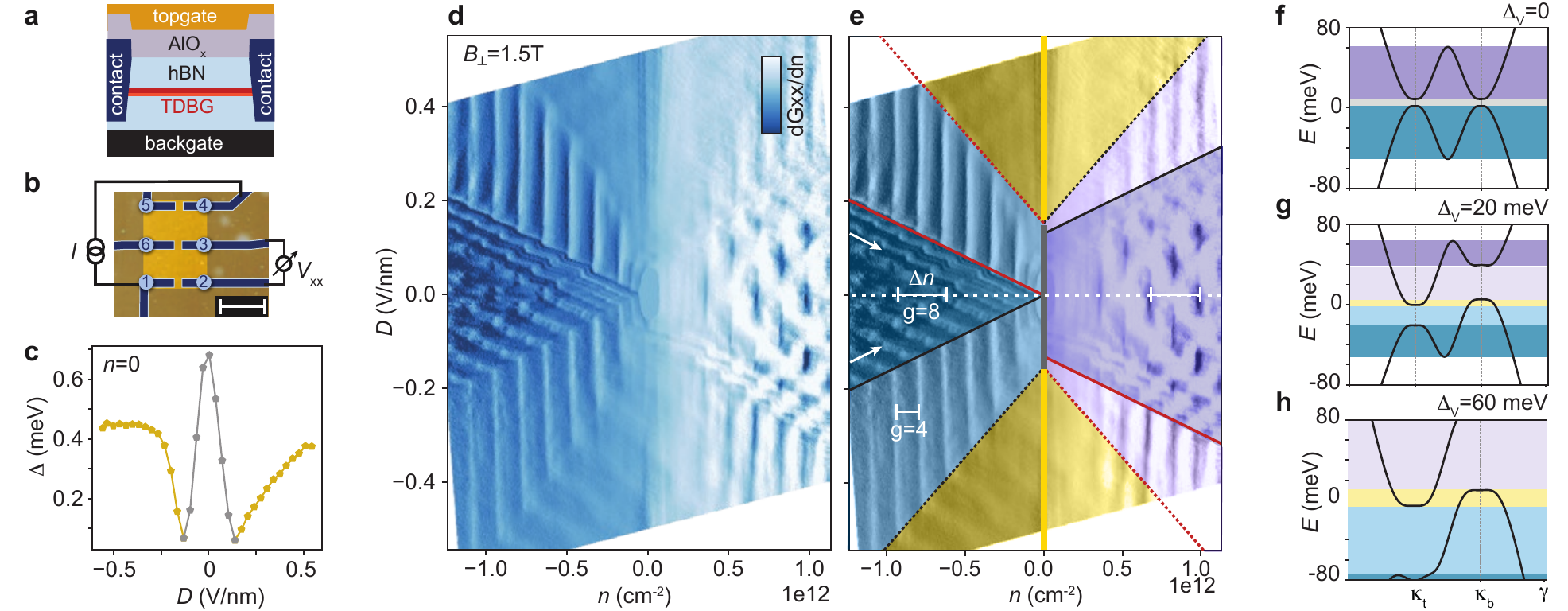}
	\caption{\textbf{Establishing electron-hole fluids}. a) Side view and
	b) top view of the stack. $\Rxx \equiv V_\mathrm{23}/I_\mathrm{14}$. Scale-bar: $\SI{2}{\mu m}$. c) From thermal activation measurements we obtain $\Delta(D)$. 
	d) $\partial/\partial n\,(1/\Rxx)(D,n)$ at $\Bperp=\SI{1.5}{T}$ and $T=\SI{1.2}{K}$ exhibits SdH oscillations that distinguish the regions highlighted in e).
	In the yellow region, electrons and holes coexist.
	f)-h) Single-particle band structures for different $\DeltaV\propto D$.
	}\label{fig:2}
\end{figure}

We now discuss our device in more detail by showing a side view schematic in Fig.~\ref{fig:2}a. We tune the density $n=(\Ct\Vt+\Cb\Vb)/e$ and displacement field $D=(\Cb\Vb-\Ct\Vt)/2\epsilon_0$ by applying voltages $\Vt$ and $\Vb$ to top and back gate electrodes. Here, $\Ct,\Cb$ are the corresponding capacitances per unit area. The resistance peak in Fig.~\ref{fig:1}c is measured using the contact geometry in Fig.~\ref{fig:2}b. 
From the decrease of the resistance peak with increasing temperature we extract a gap $\Delta$ using 
the Arrhenius law $R_{xx} \propto \text{exp}(\Delta/2k_B T)$. The extracted dependence of $\Delta$ on $D$ at total denstiy $n=0$ is shown in \ref{fig:2}c. The data reveals a gap around $D=0$ (gray line) that closes with increasing $|D|$. Another gap is opened at large $|D|$ (yellow line). 

In order to identify the conditions for the formation of the gap we present measurements of Shubnikov-de Haas (SdH) oscillations for different $n$ and $D$. In Figs.~\ref{fig:2}de we plot the numerical derivative of the inverse resistance $\partial/\partial n\,(1/\Rxx)$ (this quantity is chosen for best visibility of relevant features) measured in a perpendicular magnetic field $\Bperp=\SI{1.5}{T}$. Along the dashed line at $D=0$ in Fig.~\ref{fig:2}e, the spacing between SdH oscillations corresponds to a band degeneracy $g=8$. The degeneracy is lifted by changing $D$ and SdH lines with two slopes emerge (indicated with arrows) in the dark blue and the dark purple region. 
In the light blue and light purple region, the pattern changes and lines with $g=4$ that are parallel to the $n=0$ line are seen. The slope of the SdH lines then changes again in the yellow region, and the oscillations become weaker.

These observations suggest the presence of two subbands in the dark regions. Their energy offset is tuned by $D$, thus the subbands are related to the top and bottom bilayer \cite{Rickhaus2019,DeVries2020}. In the light blue/ light purple region, only one subband exists and SdH are independent of $D$. Since the displacement field changes the energetic offset of the subbands, the phase at large $D$ and small $n$ (yellow region), which contains two subbands, corresponds to an electron-hole fluid, in agreement with the slopes in the Landau fan diagram Fig.~\ref{fig:1}d. Importantly, the observed gap at $n=0$ and large $D$ occurs in the midst of this two-subband region, thus strongly suggesting that it originates from charge carrier correlations.
The region boundaries are well described using an electrostatic model based on parabolic subbands (see SI section \ref{sec:electrostatics} and \ref{sec:asymmetry}) where we also discuss the observed asymmetry with respect to $n$.

The experimental findings are in line with the single-particle band structures shown in Fig.~\ref{fig:2}f-h at $\twisttheory$ for different adjacent layer on-site energy differences $\DeltaV$ (for conversion of $D$ to $\Delta_V$ and the slight difference in $\twist$ between theory and experiment see SI, section \ref{supp:Fabrication}). 
At $\DeltaV=0$ (Fig.~\ref{fig:2}f) two subbands of the four-fold spin/valley degenerate bilayer graphene emerge near the $\Kt$ and $\Kb$ points, leading to a degeneracy $g=8$. Changing $\DeltaV$ (or $D$) breaks the layer degeneracy. With $\DeltaV$, single-band regions with $g=4$ emerge (Fig.~\ref{fig:2}gh). Around charge neutrality, bands with opposite carrier type coexist and the band gap is closed. Regarding the two gaps that we observe in Fig.\ref{fig:2}c, only the gap at $n=0$ ($E=0$) and $\DeltaV=0$ is captured by the single-particle band structure (Fig.~\ref{fig:2}f). This gap originates from a combination of crystal fields, which lead to electron transfer from the outer to the inner layers \cite{Rickhaus2019b} and localized states in the moiré lattice \cite{Culchac2020}. The absence of a gap at $n=0$ and finite $\DeltaV$ in the single-particle band structure agrees with our previous suggestion that the experimental gap in this region emerges from electron-hole correlations.

\begin{figure}
\centering
\includegraphics[width=0.5\textwidth]{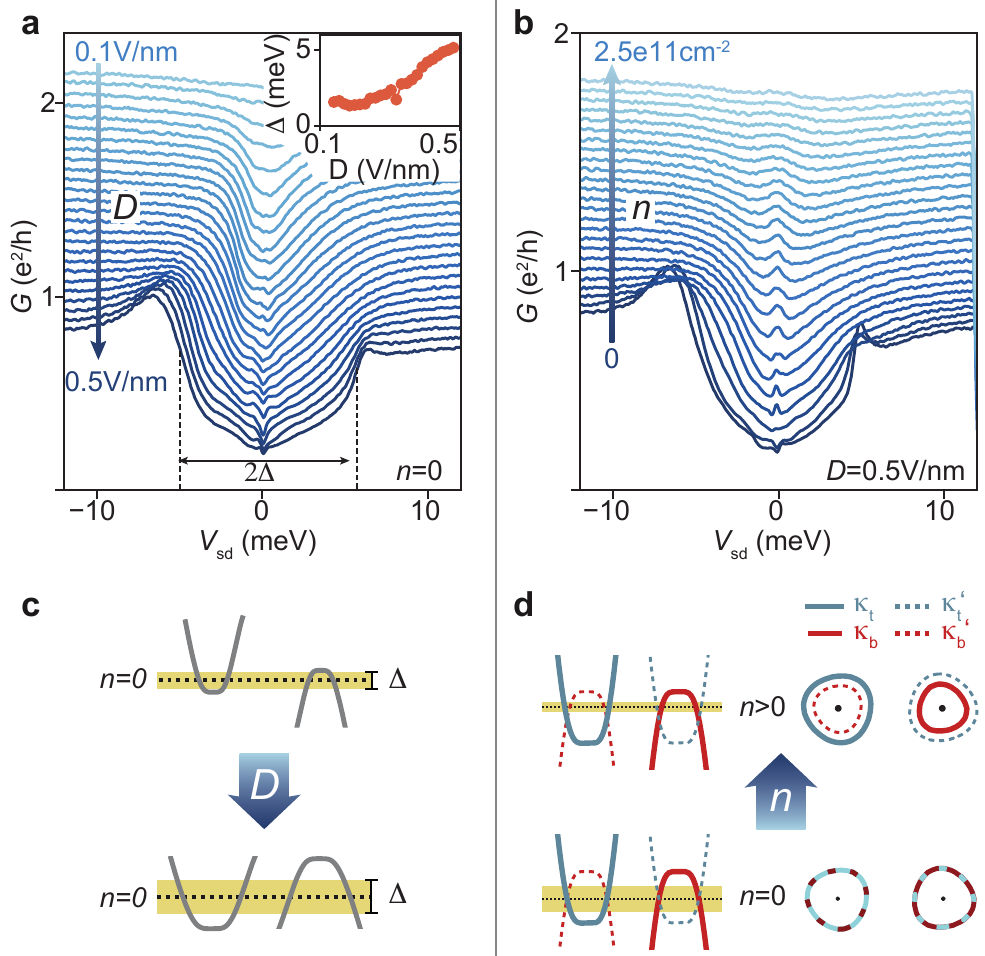}
	\caption{\textbf{Nesting}. 
	a) $G(\Vsd,D)$ at $T=\SI{0.1}{K}$ measurements reveals the appearance of a gap $\Delta$.  Curves are offset from the $\Dmax$ trace. The inset shows extracted $\Delta(D)$, revealing that $\Delta$ increases with $D$.
	b) $G(\Vsd, n)$ at $\Dmax$ showing that the gap vanishes with increasing $n$, where Fermi surface nesting falters. 
    c) The size of Fermi surfaces at $n=0$ and masses $\meff$ increase
	with $D$.
	d) Depicted is the calculated dispersion and Fermi surface around $\Kt$ (blue) and $\Kb$ (red) in the $K$ (solid) and $K'$ (dashed) valley. For details, see SI section \ref{sec:valley}. 
}\label{fig:3}
\end{figure}

We now discuss the dependence of the correlated gap on the parameters $D$ and $n$ which change the sizes of the electron-hole Fermi surfaces. We show the two-terminal conductance as a function of DC bias $G(\Vsd)$ in Fig.~\ref{fig:3}a and observe coherence peaks at $n=0$ and large $D$, suggesting the formation of a coherent ground state. In agreement with thermal activation measurements, the gap size $\Delta$ increases with $D$.
At $\Dmax$ and $n=0$, $\Delta\approx\SI{5}{meV}$ ($\Delta$ is determined by the inflection points). We note that usually, thermal activation energy and the bias gap agree, but differ by an order of magnitude here. We speculate that this originates from complex thermal breakdown of the correlated gap and estimate that the bias measurement overestimates $\Delta$ by $\sim10\%$ by additional series resistances (see SI section \ref{supp:Fabrication}).
In Fig.~\ref{fig:3}b we show the evolution of $\Delta(n)$ at $\Dmax$. Upon increasing $n$, the gap smears out and vanishes at $n\approx\SI{2.5e11}{cm^{-2}}$. 

The correlated electron-hole gap vanishes with decreasing $|D|$ and increasing $|n|$. The first effect indicates the importance of effective mass and Fermi surface size, both of which increase with increasing $|D|$, as sketched in Fig.~\ref{fig:3}c. The $|n|$ dependence, on the other hand, suggests that an increasing asymmetry between electron and hole Fermi surfaces (blue and red) weakens the gap, as depicted in Fig.~\ref{fig:3}d. The observation that a correlated state emerges once Fermi surfaces match in size (at $n=0$), and that the gap falters for small asymmetry, strongly suggests that nesting of Fermi surfaces (i.e. contours in 2D) plays a crucial role for the formation of the correlated state.

\begin{figure}
\centering
\includegraphics[width=1\textwidth]{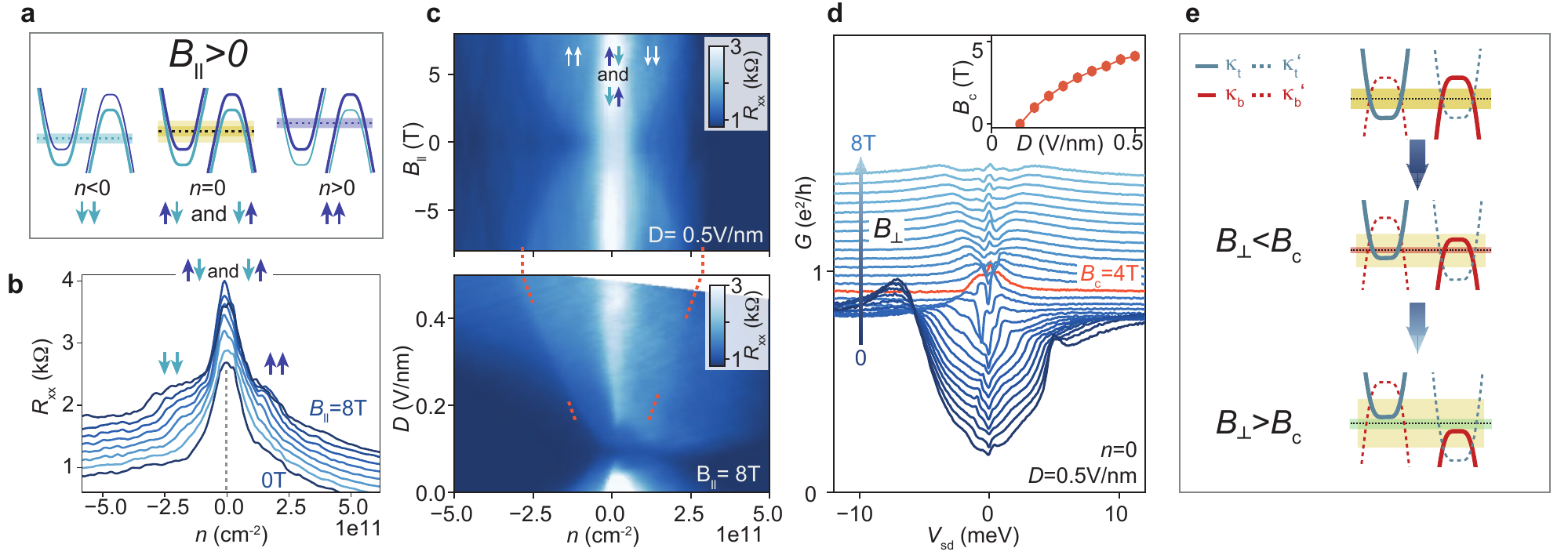}
\caption{\textbf{Spin polarized correlated gap.}
	a) Application of $\Bparallel$ splits the bands according to their magnetic moments, giving rise to regions where  only one of the split bands is gapped. The arrows indicate the magnetic moments of the charge carriers.
	b) The partially gapped regions lead to an increase in $\Rxx$ at finite $n$.
	c) $\Rxx(\Bparallel,n,\Dmax)$ and $\Rxx(D,n,\Bparallel=\SI{8}{T})$ maps. The orange lines are calculated using $\Delta(D)$ and $m^*$ from the experiment, without fitting parameters.
	d) The gap is closed by $\Bperp$. In the inset we plot the critical field, $\Bc$, as a function of $D$.
	e) The microscopic valleys K (solid) and K' (dashed) shift in energy with $\Bperp$, leading to same-valley pairing at $n=0$. For simplicity, Zeeman splitting is not sketched. All measurements in the figure are taken at $T=\SI{0.1}{K}$.
}\label{fig:4}
\end{figure}

We can probe this hypothesis by shifting the Fermi surfaces using a magnetic field. The idea of such a measurement in parallel field is depicted in Fig.~\ref{fig:4}a. At finite $\Bparallel$, the bands shift by the Zeeman energy. Now, nesting (equal Fermi surface) between \textit{opposite} magnetic moments is possible at $n=0$, whereas nesting between \textit{same} magnetic moments is possible at $n\neq0$. Indeed, the measured $\Rxx(n,\Bparallel)$ traces are consistent with this concept (Fig.~\ref{fig:4}b). At $\Bparallel=\SI{8}{T}$, a shoulder in resistance at finite $n$ is visible, agreeing with a partially gapped bandstructure where half of the carriers form a correlated state while the other magnetic moment-bands remain conducting. The width of the shoulder (in $n$) increases with the Zeeman energy.
How the region of enhanced resistance changes as a function of $\vert\Bparallel\vert$ and $D$ can be observed in Fig.~\ref{fig:4}c. We can model the dependence on Zeeman energy by a basic model giving the dashed lines in Fig.~\ref{fig:4}c (for details see SI section \ref{sec:zeemann} and \ref{sec:gapinBparallel}).

The behavior in perpendicular magnetic  field $\Bperp$ is more complex.
In Fig.~\ref{fig:4}d we show $G(\Vsd, \Dmax)$ traces for different $\Bperp$. We see that the gap is closed at a critical field $\Bc=\SI{4}{T}$ (orange trace) whose $\Bc(D)$-dependence is plotted in the inset. In a semi-classical picture, the bands shift with $\Bperp$ due to the valley-Zeeman effect. The valley g-factor in bilayer graphene $\gv\sim20-120$ \cite{Yongjin2019} has opposite signs in conduction and valence bands (see SI \ref{sec:LLs} for details). $\Bperp$ therefore increases both electron and hole densities in one valley and decreases both in the other valley, as illustrated in Fig.~\ref{fig:4}e . If the correlated e-h gap is formed by wavefunction nesting, the gap in one valley (dashed bands - yellow gap) increases with $\Bperp$ and decreases in the other valley (solid bands - red gap). Thus, the completely gapped part of the bandstructure is decreasing with $\Bperp$, in agreement with the observations. 
In this picture, the gap opening for $\Bperp>\Bc$ (green) can be interpreted as a single-particle gap at $K'$ (green in Fig.~\ref{fig:4}e) and a correlated gap at $K$, {\it i.e.} as the formation of a valley polarized correlated state.

The observations in magnetic field are entirely consistent with basic models that assume Fermi surface nesting. We now confirm nesting of electron-hole surfaces by HF calculations where we incorporate the Coulomb potential and calculate the correlated gap $\Delta$ self-consistently (see SI section \ref{sec:HF}). Note that the moiré potential is not relevant for the emerging correlations since tunneling between top and bottom bilayers is weak at the intermediate twist in the relevant energy range. From HF, we obtain the correlated bands as shown in Fig.~\ref{fig:5}ab. A gap is opened by electron-hole correlation, and linearly increases with adjacent layer on-site energy difference $\Delta_V$ for $\Delta_V \leq 25$ meV.  Both the gap size and the linearity with respect to displacement field agree with experiments, as shown in Fig.~\ref{fig:5}c. We point out that this behavior is a result of decreasing static dielectric constant $\epsilon(\mathbf{q})$ with $D$ (see SI section \ref{sec:dielectricconstant}). For $\Delta_V > 25$ meV, the gap decreases as a result of the non-negligible moir\'e band asymmetry (seen in Fig.~\ref{fig:5}c). 

We note that Fermi surfaces in the same valley (e.g. $\Kt$ and $\Kb$) are asymmetric even at $n=0$ as they are not perfectly circular due to interlayer tunneling between the middle two layers. However, the Fermi surfaces of opposite valleys (e.g. $\Kt$ and $\Kbprime$) match, suggesting that the correlated state is formed out of charge carriers from opposite valleys at $n=0$. We show the calculated Fermi surfaces in SI \ref{sec:valley}.

Our model allows us to adjust the interlayer hopping parameter $\gamma_1$ between dimer sites in each bilayer in order to change $m^*$. 
The effective mass $m^*$ generally depends on parameters ($w$, $\theta$, $\Delta_V$, $\gamma_1$, $v_\text{F}$). If $w \neq 0$, changing $w$, $\theta$ or $\Delta_V$ will change both $m^*$ and layer polarization Prob($\kappa_{\text{top}}$). If $w=0$, changing $\theta$ or $\Delta_V$ have no effect on $m^*$, while changing $\gamma_1$ or $v_\text{F}$ tunes $m^*$ without any effect on layer polarizations. In order to investigate the effect of $m^*$ solely on the correlated gap, we adjust the interlayer hopping parameter $\gamma_1$ (Eq.(\ref{Eq_TBernal}) in SI \ref{sec:model_H}) between the dimer sites in each bilayer.  We observe an almost linear increase $\Delta(m^*)$ (see Fig.~\ref{fig:5}h), supporting our previous statement that $\meff$ determines the correlation strength (Fig.~\ref{fig:1}b). Theory does not show a strong dependence of pairing on the momentum difference between Fermi surfaces, only on the effective mass. This is in agreement with the excitonic character of a DW. In summary, the gap opening in HF calculations suggests that the correlated electron-hole state is best described as a DW.

In conclusion, we have revealed the appearance of a correlated gap in TDBG in the intermediate twist angle regime, formed out of electrons and holes with equal Fermi surface.
The wavefunctions in the top/bottom bilayer can be tuned individually making it possible to enter a region where electron and hole bands coexist. In this region, we observe a gap that likely emerges from nesting of electron and hole Fermi surfaces. The spin and valley dependence of electron-hole pairing are both strongly influenced by the application of a magnetic field. The correlated electron-hole state can be viewed as an excitonic insulator and is expected to exhibit counterflow superfluidity.
Quantum phase transitions between DWs and disordered states can be controlled without chemical doping
by varying $n$, or by varying $D$ at $n=0$, and such tunability could  provide a new window on non-Fermi-liquid physics.



\section*{Acknowledgements}
We acknowledge financial support from the European Graphene Flagship, the Swiss National Science Foundation via NCCR Quantum Science. P. Rickhaus acknowledges financial support from the ETH Fellowship program. Growth of hexagonal boron nitride crystals was supported by the Elemental Strategy Initiative conducted by MEXT, Japan and the CREST (JPMJCR15F3), JST. AHM and JZ were supported by the National Science Foundation through the Center for Dynamics and Control of Materials,  an NSF MRSEC under Co- operative Agreement No. DMR-1720595 and by the Welch Foundation under grant TBF1473.
\\
Data and materials availability: All data and code is available under Ref.\cite{Rickhaus2021_datarep}

\clearpage
\setcounter{figure}{0}
\setcounter{page}{0}
\renewcommand{\thefigure}{S\arabic{figure}}
\renewcommand{\theequation}{S\arabic{equation}}

\subsection{Materials and Methods}\label{supp:Fabrication}

We fabricate encapsulated TDBG by the tear-and-stack method \cite{Kim2016}. Twisted double bilayer graphene is encapsulated in a top/bottom hBN with thicknesses $\SI{27.0/50.3}{nm}$ and features a graphite back gate\cite{Zibrov2017a}. Clean device areas are identified by atomic force microscopy. TDBG is contacted by edge contacts \cite{Wang2013} (blue in the schematic in Fig.~\ref{fig:2}a). After deposition of local top gates (not shown in the schematic), the TDBG is etched (brown in the top-view in Fig.~\ref{fig:2}b) and $\SI{30}{nm}$ of AlO$_x$ are deposited. Finally, we evaporate a global top gate (yellow). Throughout the measurements, the local and global top gates are tuned such that a uniform potential forms, therefore we will not discuss the effect of the local gate.

We measure $\Rxx$ by passing an AC current between contacts 1 and 4 (Fig.~\ref{fig:2}b), and measuring the voltage $V_\mathrm{xx}$ between contacts 2 and 3. For the finite bias measurements in Fig.~\ref{fig:3} we apply AC+DC voltage between contacts 3 and 6 and measure the resulting AC current to obtain the differential conductance $G(\Vsd)$. We have measured all other devices (3x6 contacts) on the stack and obtained comparable results, shown in section \ref{sec:otherdevice}. All measurements are performed at a temperature of $\SI{100}{mK}$, unless stated otherwise.

 For the measurement of $\Delta(\Vsd)$, an additional series resistance $R_s$ has  to be considered. The measurement of $R$ at large densities gives an estimate for $R_s\approx\SI{2}{k\Omega}$ which is an order of magnitude smaller than the resistance in the gap, indicating that $\Delta$ can be estimated well from $G(\Vsd)$. 

The external field $D=1/2\epsilon_0(\Cb\Vb-\Ct\Vt)$ is simulated in the band structure calculation by an on-site energy difference $\DeltaV$ between adjacent layers, where the on-site energy from the top-most to the bottom-most layer is $U=\DeltaV(-3/2,-1/2,1/2,3/2)$. The relation of $\DeltaV$ to the external field $D$ for bilayer is $\DeltaV=eDd/\epsilon =eD\epsilon_0/\Cm$, with $\Cm\approx\SI{7.5}{\mu F/cm^2}$ the capacitance between the graphene layers \cite{Rickhaus2019}. Therefore, for the four layers, $\DeltaV=bD$ with $b=e\epsilon_0 /2\Cm\approx\SI{59}{meV/(Vnm^{-1})}$. We would like to note that this conversion has to be taken with caution and that a more reliable conversion can only be obtained by self-consistently calculating the layer on-site energies in an external field.

 For our intermediate twist angle $\twist$, the density of full filling of the first band ($\ns\approx\SI{13e12}{cm^{-2}}$) is outside the measurement range. However, we have three possibilities to determine the twist angle. First, we do observe Landau levels emerging from the band edge, allowing us to estimate $\ns$. Second, the Lifshitz transition is clearly visible in our devices and its critical density is characteristic for a certain twist angle. Finally, the Hofstadter butterfly pattern, where we clearly observe flux quanta through up to 37 moiré unit cells, allows us to determine the size of the unit cell and therefore the twist.
All three methods are explained in Ref.\cite{DeVries2020}. We find a twist angle of $2.37^\circ$ with little variation ($\pm0.04^\circ$) along the $\SI{15}{\mu m}$ long device. 

\begin{figure}[b]
	\centering
	\includegraphics[width=1\textwidth]{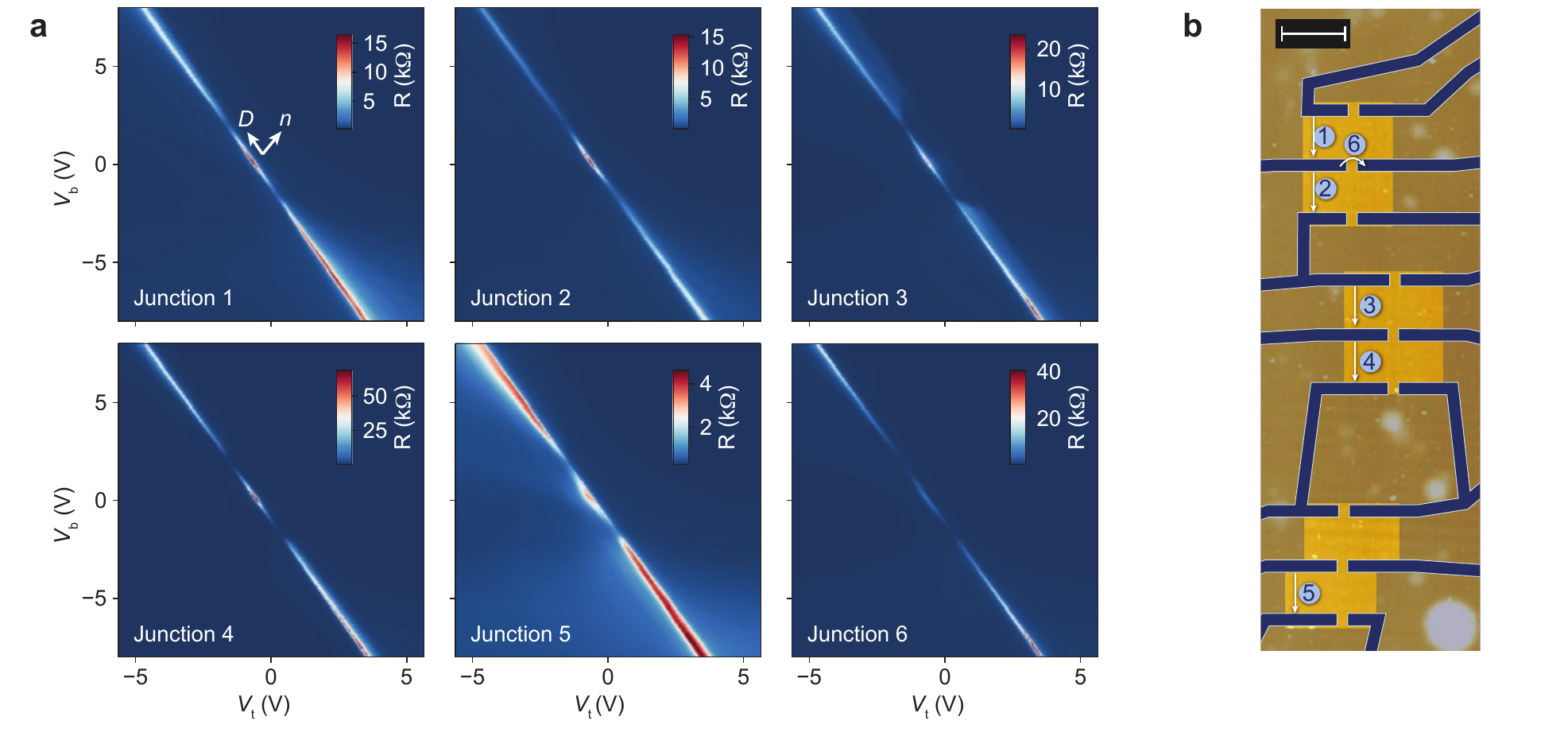}
	\caption{a) Two-terminal resistance $R$ as a function of top- and back gate voltage ($\Vt$, $\Vb$). Direction of $D$ and $n$ are indicated.
		b) Device schematics, contacts blue, etched parts brown. The junctions measured in a) are indicated with numbers. Scale-bar: $\SI{2}{\mu m}$.
	}\label{suppfig:devices}
\end{figure}

In the main-text we use $\twisttheory$ for the band structure calculations, but the experimentally determined angle is $\twist$. This variation has little influence on our interpretation, as the band structure is rather robust against small changes in $\theta$ in the intermediate twist angle regime.

\subsection{Measurements in other devices}\label{sec:otherdevice}

In Fig.~\ref{suppfig:devices} we show two terminal resistance measurements $R(\Vt,\Vb)$ for different junctions on the device. A resistance peak at finite $D$ and $n=0$ is observed for all devices, i.e. all devices exhibit the correlated electron-hole gap.

\subsection{Density wave state at $\theta=1.98^\circ$}\label{sec:twist2}
\begin{figure}[ht]
	\centering
	\includegraphics[width=1\textwidth]{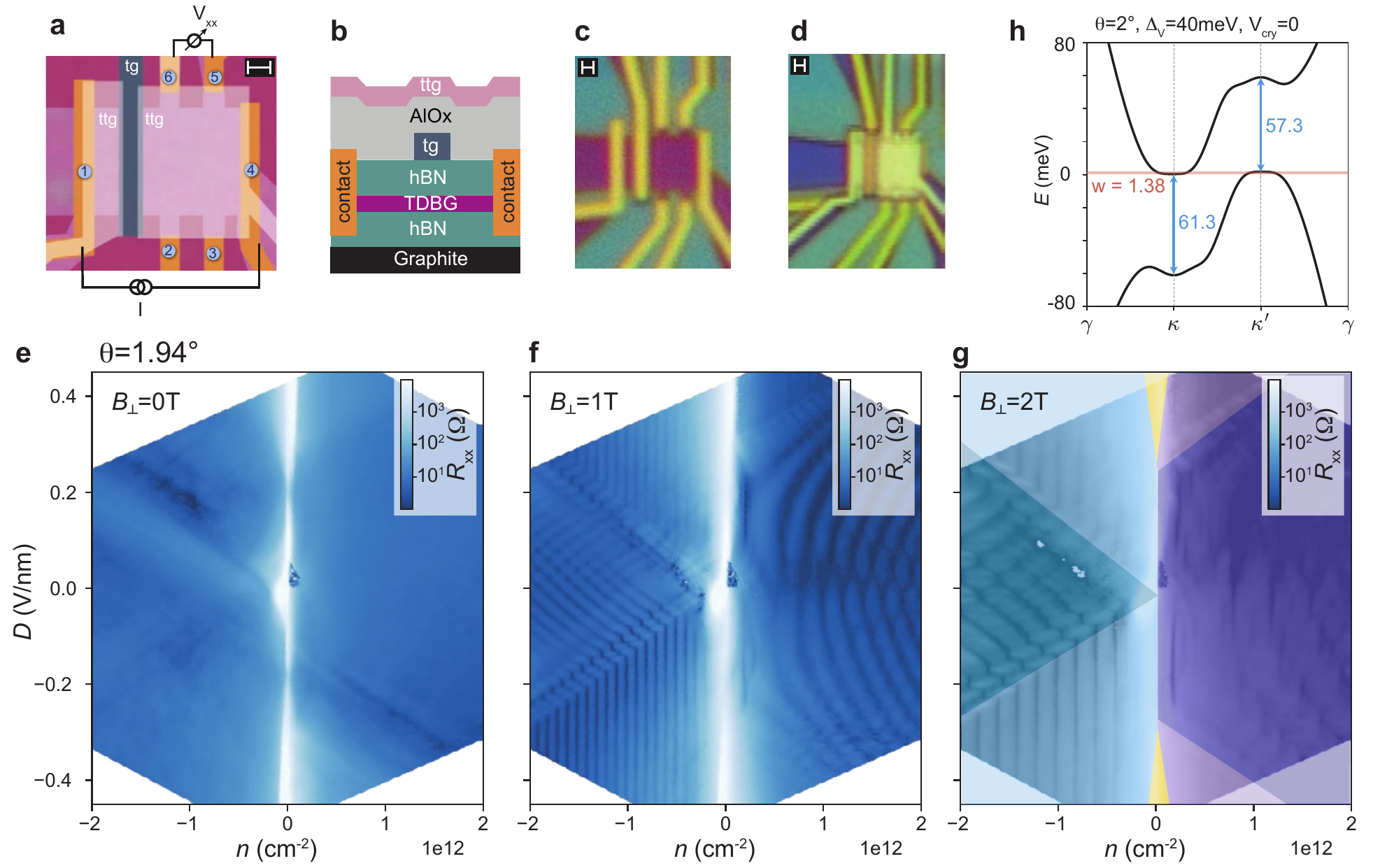}
	\caption{
	a) Schematic top-view of the device with a twist-angle of $1.98^\circ$. Current is passed from contact 1 to 4 while the voltage is measured between contacts 5 and 6. The device has a topgate (tg) and a top-topgate (ttg) separated by a layer of aluminium-oxide, as shown in (b). For all the measurements, tg and ttg are tuned such that a uniform density $n$ is achieved.
	c) Optical image before and after (d) deposition of the ttg. Scale-bar: $\SI{500}{nm}$.
	e)-g) $\Rxx(n,D)$ for different magnetic fields. In g) we color the single- and double-band regions for comparison with Fig.\ref{fig:2}e.
	h) Corresponding single-particle band structure for $\theta=2^\circ$.
	}\label{suppfig:twist2}
\end{figure}

\begin{figure}[ht]
	\centering
	\includegraphics[width=1\textwidth]{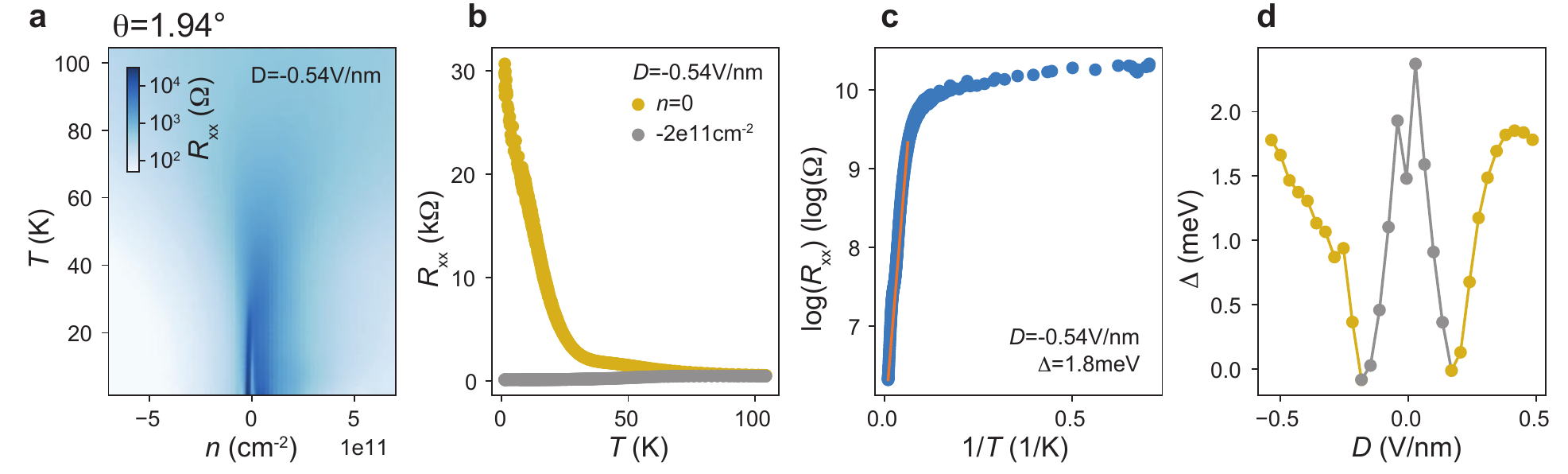}
	\caption{a) $\Rxx(n,T)$ at $D=\SI{-0.54}{V/nm}$.
	b) The resistance peak at $n=0$ and resistance at a small offset density $n=\SI{-2e11}{cm^{-2}}$ as a function of $T$.
	c) Fitting the Arrhenius law.
	d) Extracted $\Delta(D)$.
	}\label{suppfig:twist2-thermal}
\end{figure}

In Fig.~\ref{suppfig:twist2} and \ref{suppfig:twist2-thermal} we show data of another heterostructure with a twist of $\theta=1.98^{\circ}$. The value of $\theta$ has been determined from Brown-Zak oscillations (see last paragraph in section \ref{supp:Fabrication}). The device is shaped in a Hall-bar geometry (Fig.
~\ref{suppfig:twist2}a-d) with a local top- (tg) and a global toptop gate (ttg). During all the measurements, tg is tuned such that the region below both gates is at the same density $n$. 
In Fig.\ref{suppfig:twist2}e-g we show $\Rxx(n,D)$. The pattern of Shubnikov-de-Haas oscillations exhibit a similar behavior to the device in the main text at $\twist$, i.e. at densities lower than the Lifshitz transition ($n_\mathrm{Lifshitz}\approx\pm\SI{3.2e12}{cm^{-2}}$) the system can be described by weakly coupled bilayer graphene bands stemming from the $\kappa$ and $\kappa'$ points which can be shifted in energy by $D$. The band overlap at $n=0$ and large $\Delta_V$ is smaller than in the previous case, therefore we could not confirm the coexistance of electron and hole bands at $n=0$ and large $D$ by Shubnikov-de-Haas oscillations experimentally (as we did in Fig.
~\ref{fig:1}d for $\twist$). However, the single-particle electron and hole bands (Fig.~\ref{suppfig:twist2})h overlap, $w=\SI{1.38}{meV}$. This overlap is significantly smaller than the respective size at $\theta=2.3^\circ$ ($\SI{10.9}{meV}$, Fig.~\ref{suppfig:asymmetry}).
Nevertheless, the single-particle band-structure is not gapped but a thermally activated gap at large $D$ and $n=0$, is observed in transport. This correlated gap stems from a density wave state. The dependence of $\Rxx(n,T)$ at $D=\SI{-0.54}{V/nm}$ is shown in Fig.~\ref{suppfig:twist2-thermal}a and analyzed in Fig.~\ref{suppfig:twist2-thermal}b and Fig.~\ref{suppfig:twist2-thermal}c. The resistance strongly decreases with temperature, from a maximal value of $\SI{30}{k\Omega}$ at $T=\SI{1.2}{K}$ to $\SI{600}{\Omega}$ at $\SI{50}{K}$. Such behavior is absent at a small offset density of $n=\SI{-2e11}{cm^{-2}}$. By fitting the Arrhenius law (Fig.
~\ref{suppfig:twist2-thermal}c) we extract a gap of $\Delta=\SI{1.8}{meV}$, which is significantly larger than the gap we extracted from thermal activation at $\twist$ and $D=\SI{0.5}{V/nm}$, i.e. $\Delta=\SI{0.4}{meV}$. The trend can be explained by the increased effective mass at smaller $\theta$. We also plot $\Delta(D)$ in Fig.~\ref{suppfig:twist2-thermal}d. 

\subsection{Measurement at further twist angles}\label{sec:othertwist}
\begin{figure}[ht]
	\centering
	\includegraphics[width=1\textwidth]{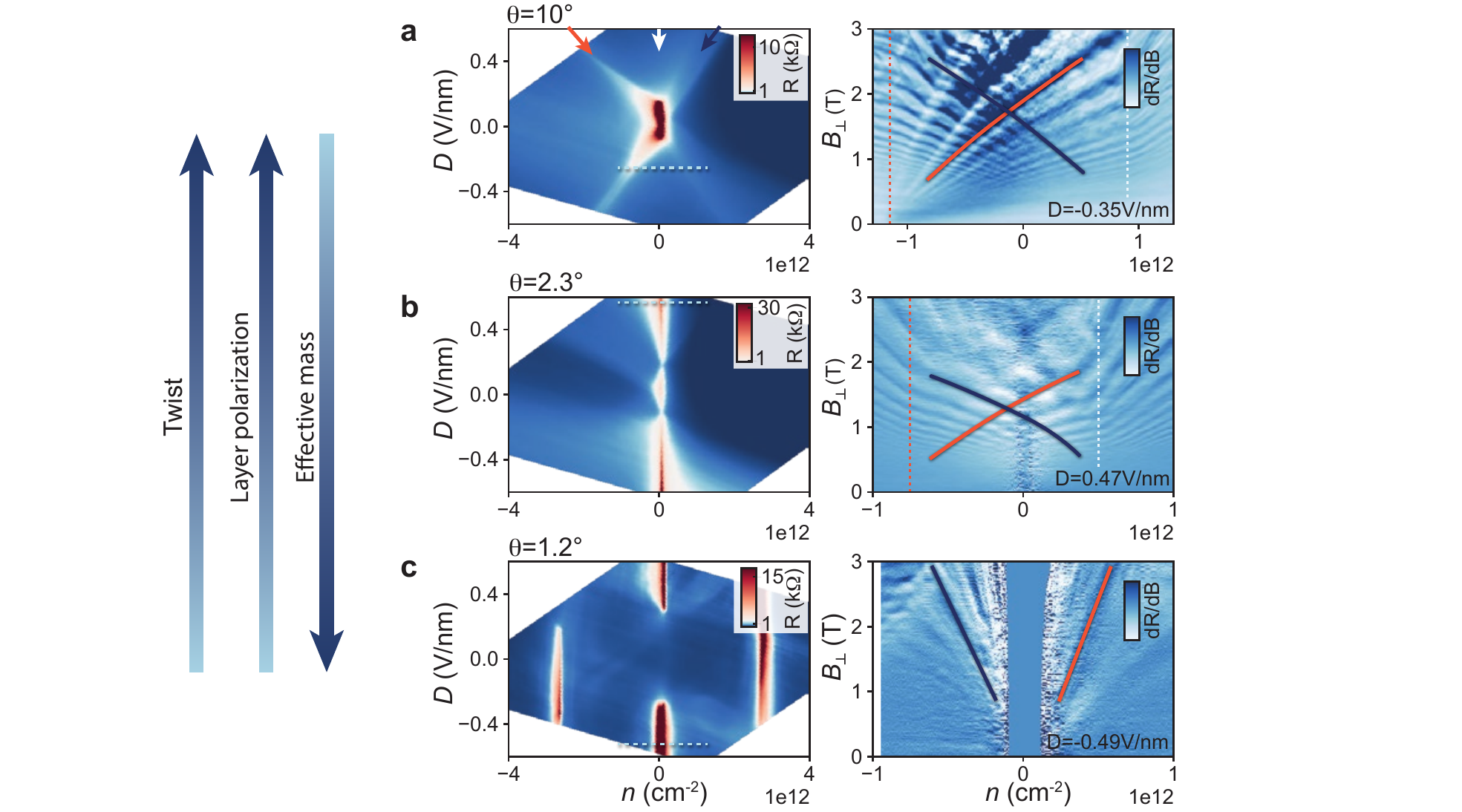}
	\caption{Comparison of $R(n,D,B=0)$ and Landau fans at finite $D$ for different twist angles. For better visibility, the numerical derivative $dR/dB$ of the Landau fan is shown. 
	a) The large-twist angle device ($\theta=10^\circ$) exhibits a strong resistance peak around $(n,D)\sim(0,0)$ due to crystal fields. At finite $D$ and $n=0$, the system is not gapped, even though electron and hole bands coexist as seen in the Landau fan, where electron-hole resonances are indicated with red/blue lines. They originate from the band edges, indicated with dashed lines.
	b) electron-hole resonances are also observed in the Landau fans for $\twist$. However, for $|D|>\SI{0.1}{V/nm}$ and $n=0$, a gap is observed, as discussed in the previous figures.
	c) The device with a smaller twist of $\theta=1.2^\circ$ exhibits a fundamentally different behavior. No gap is observed at  $(n,D)\sim(0,0)$, but at $(|n|,D)\sim(\SI{3e12}{cm^{-2}},0)$ (full filling of the moiré unit cell) and for $(n,|D|)\sim(0,>\SI{0.3}{V/nm})$ which is due to a band-gap in the band structure at finite $D$. In this case, the Landau fan emerges from $n=0$ and electron-hole resonances do not cross. 
	}\label{suppfig:twists}
\end{figure}
Here we analyze the impact of the twist angle by contrasting $\Rxx(n,D)$ and Landau fans at $\twist$ to measurements at $\theta=10^\circ$ and $\theta=1.2^\circ$. 
We first consider the large twist device, where we observe decoupled behavior. In the $R(n,D)$ map shown in Fig.~\ref{suppfig:twists}a, a large resistance peak around $(n,D)=(0,0)$ is measured that is attributed to the presence of crystal fields\cite{Rickhaus2019b}. By increasing the displacement field, two features of increased resistance split up (marked with orange and blue arrows), corresponding to charge neutrality lines in the top and bottom bilayer. Importantly, around $n=0$ no resistance peak is observed (white arrow), in agreement with the observations in Ref.\cite{Rickhaus2019b}. The Landau fan at finite $D=\SI{-0.35}{V/nm}$ in Fig.~\ref{suppfig:twists}b agrees with the expected coexistence of electron and hole bands.
At large twist angles, it is therefore possible to observe the coexistence of electron and hole Fermi surfaces, but these do not form a correlated state. 

Decoupled behavior and overlapping electron-hole bands are also observed for the $\twist$ device, Fig.~\ref{fig:5}b. In contrast to $\theta=10^\circ$, the single-band regions occupy a larger density range due to increased effective mass. However, the devices can be described by a similar electrostatic model. The important difference is the occurrence of a gap at $n=0$ and finite $D$.

The small twist-angle device (Fig.~\ref{fig:5}c) exhibits a fundamentally different behavior (see also \cite{Liu2019,Shen2019, Burg2019}). Resistance peaks are observed at $n\approx\SI{\pm3e12}{cm^{-2}}$ due to filling of the first band. At $D=0$, there is no band gap at zero energy due to band overlap. The band overlap is removed by applying a displacement field $\vert D\vert>\SI{0.3}{V/nm}$, where a single particle gap appears. The Landau fan in this regime reveals that electron and hole bands do not overlap at finite $D$, they rather emerge from the gap around $n=0$. This is in agreement with band structure calculations that exhibit a single-particle gap at finite $D$ \cite{Burg2019,Li2019}. At small twist angles it is therefore not possible to engineer a coexistence of electron and hole Fermi surfaces by increasing the displacement field. This is due to the fact that the wavefunction is no longer layer polarized, therefore it is not possible to charge the upper bilayer with electrons and the lower bilayer with holes.

The measurements strongly suggest that an intermediate twist angle is important. We argue that, on the one hand, the twist needs to be large enough such that layer-polarized, bilayer bands emerge from the $\Kt$,$\Kb$ points, see also \cite{DeVries2020}. The bilayer bands can be individually controlled by top/back gate electrodes. In strong contrast to correlated states that are formed out of flat bands, here, single-particle band structure calculations reveal the states out of which a correlated groundstate is formed. On the other hand, the twist angle needs to be small enough such that the effective mass of the bands is sufficiently large to obtain a well observable gap, as we argue in the main text.

\subsection{Thermal activation of the gap and asymmetry of the gap with respect to $D$}\label{sec:thermal}
\begin{figure}[ht]
	\centering
	\includegraphics[width=1\textwidth]{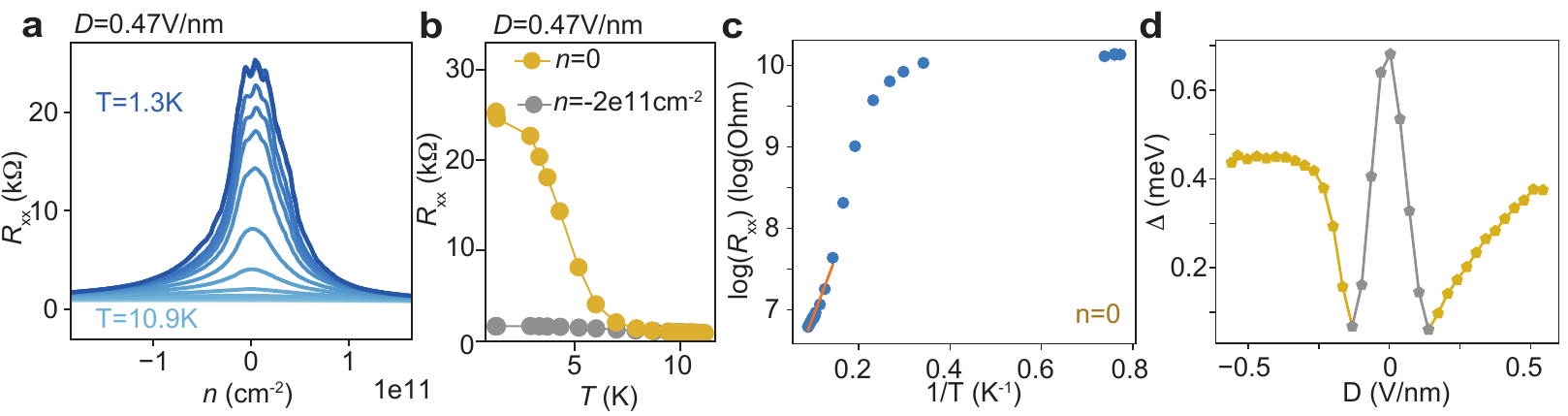}
	\caption{Additional data on the thermal activation of the gap. a,b) The peak in $\Rxx$, caused by the correlated gap, is vanishing quickly with temperature.
		c) Using a linear fit (orange) in the Arrhenius plot allows to extract the gap size $\Delta$. 
		d) Obtained $\Delta$ for different $D$, as in the main text.
	}\label{suppfig-T}
\end{figure}

In Fig.~\ref{suppfig-T} we show additional data on the thermal activation of the gap.

The gap at finite $D$ is asymmetric with respect to the sign of $D$. Such an asymmetry is often present in experiments with bilayer  or twisted double bilayer graphene. This can be seen in the devices at different twist angles which we present in Fig.\ref{suppfig:twists}, independent of twist angle. Presumably, it originates from an asymmetry between the outer two layers. There might be a different amount or distribution of charged impurities in the adjacent top and bottom hBN layer, respectively, caused by the stacking process.

\subsection{Effective mass}\label{sec:meff}
\begin{figure}[ht]
	\centering
	\includegraphics[width=1\textwidth]{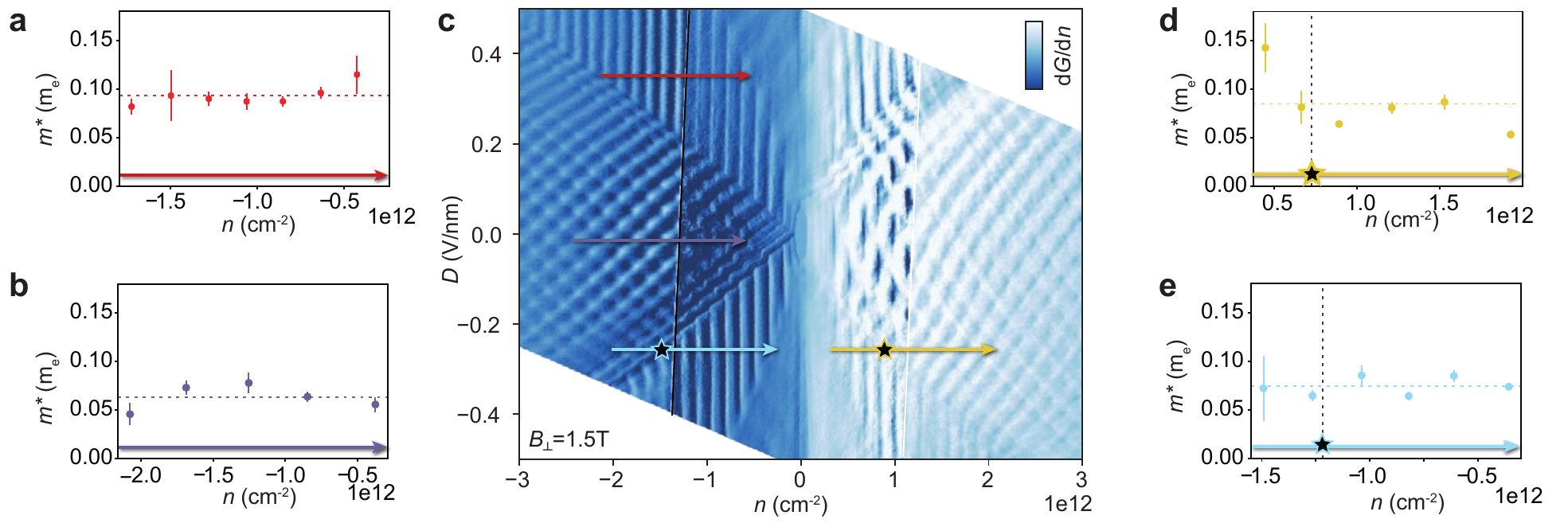}
	\caption{a)b)d)e) Extracted effective mass for different cuts shown in c), with electron mass $\me$. c) Colorscale plot of $dG/dn(n,D)$. The inner part (roughly between $\pm\SI{1e12}{cm^{-2}}$) is the same measurement as in Fig.~\ref{fig:2}d and we added a more coarse measurement for large densities.
	}\label{suppfig:meff}
\end{figure}
In Fig.~\ref{suppfig:meff} we show extracted effective masses for different cuts in the $n,D$ map for the device at $\twist$. The extraction was done by analyzing the thermal activation of Shubnikov-de Haas oscillations. The average effective mass along the red and blue cut (single valence band region) is $\meff=0.09\me$ and $0.08\me$, respectively. For the double band region we find, in average, $\meff=0.06\me$, and for the single and double counduction band region (yellow) $\meff=0.09\me$. 
\begin{figure}[ht]
	\centering
	\includegraphics[width=1\textwidth]{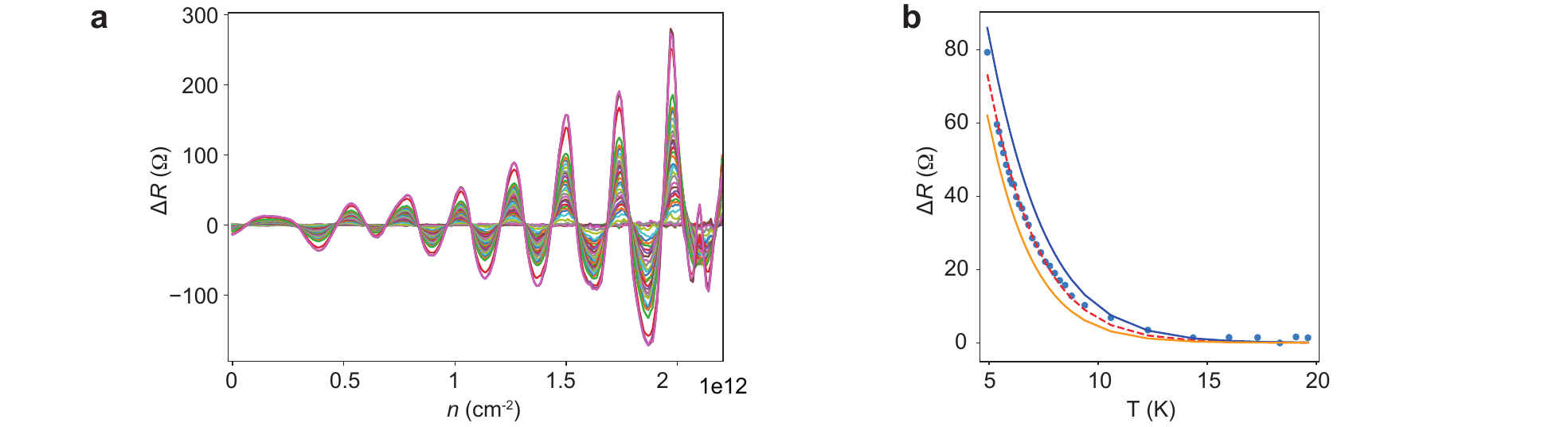}
	\caption{a) Shubnikov-de Haas oscillations with subtracted background along a cut in Fig.~\ref{suppfig:meff} for temperatures between $5$ and $\SI{20}{K}$.
		b) $\Delta R(T)$ for a fixed density in a) where $\Delta R$ exhibits a local extrema at the lowest temperature. The data points are fitted (red dashed line) to extract the effective mass. 
	}\label{suppfig:meff2}
\end{figure}

To extract the effective masses from the different regions of the $n-D$ map we remove a background from $\Rxx(n,T)$ and obtain a series of Shubnikov-de Haas (SdH) oscillations for different temperatures, as shown in Fig.~\ref{suppfig:meff2}a. We remove a polynomial background. This is done for the minimal polynomial order and the maximal polynomial order which result in a flat background, separately. For each of the two selected extremes we perform the following procedure for extracting the effective mass:
We select a peak at a given density and extract its maximum value for every different temperature. We then perform a fit of such maxima as a function of temperature using the formula 
\begin{equation}
\Delta R \propto \frac{\chi}{\sinh{\chi}} \qquad\mathrm{with:}\quad\chi = \frac{2\pi^2k_B T m^*}{\hbar e B}. 
\label{eq:fit_formula}
\end{equation}
Once obtained the value of the effective mass, $m^*$, we calculate two envelopes of equation (\ref{eq:fit_formula}) as shown in Fig.~\ref{suppfig:meff2}b. The envelopes are selected such that only one data point is outside the envelopes for $T<12.5{K}$. This is the threshold we use for plotting the error bars. We finally average the points of the two polynomial order extrema and merge the error bars such that we obtain the effective mass as a function of density with error bars taking into account both the background extraction and the fitting error, as shown in Fig.~\ref{suppfig:meff}.

\subsection{Electrostatic model on parabolic bands}\label{sec:electrostatics}
\begin{figure}[ht]
	\centering
	\includegraphics[width=1\textwidth]{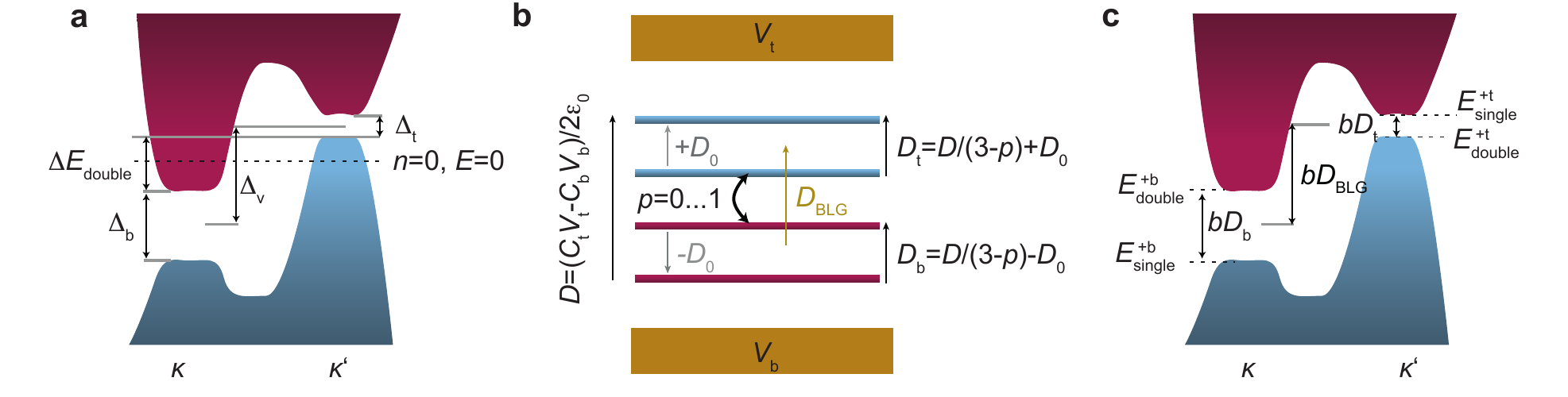}
	\caption{Important labels of the electrostatic model. a) Sketch of a band-structure at finite displacement field. Gaps in the bottom and top-layer ($\Deltat$, $\Deltab$), band-offset $\Deltabands$ and size of the electron-hole overlap region $\Edouble$ are indicated.
		b) Side-view of the stack where the density and field of the bilayer graphene layers (blue and red) is tuned by the voltages on the top gate $\Vt$ and the back gate $\Vb$. The top (bottom) bilayer is exposed to a field $\Dt$ ($\Db$), opening a gap $\Deltat$ ($\Deltab$). $\Dt$ consist of an intrinsic field $D_0$ and a contribution form the external field $D$, caused by the top/back gate voltages. $D$ is reduced by a finite tunneling probability between the inner two layers.
		c) The boundaries in Fig.~\ref{fig:2}e are calculated by determining the energies of the transitions between double and single-band region in the top (t) or bottom (b) layer. For a positive displacement field $+D$, these are $\Edouble^{+b}$, $\Edouble^{+t}$, $\Esingle^{+b}$ and $\Esingle^{+t}$.
	}\label{suppfig:model}
\end{figure}

Here we introduce a model based on parabolic bands with the goal to determine the density $n$ and displacement field $D$ of the transition between single- and double band regions. 
We start by considering the case where an external field $D$ introduces a linear charge distribution from top to bottom layer. The electric field between the top layers, between the inner layers and between the bottom layers would be $D/3$. In contrast, if charges equilibrate completely between the inner two layers, then the displacement field in the top and bottom bilayer is $D/2$. Only in this case, the electron-hole bands at the $\Kt$ and $\Kb$ point don't overlap, but touch. \\
We introduce the parameter $p$ which is the probability that charges in the inner two layers equilibrate. For $p=1$, the inner layers have the same charge density (full equilibration), for $p=0$ the charge in the inner layers is entirely determined by the external field (no equilibration). For the field between the upper two layers, we can then write:
\begin{eqnarray}
\Dt=\frac{D}{3-p}=\frac{1}{(3-p)2\eps}(\Ct\Vt-\Cb\Vb)
\end{eqnarray}
The  field between the inner layers is then
\begin{eqnarray*}
	D_{\mathrm{inner}}=D(1-p)/3
\end{eqnarray*}
For the field between the center of charges in the top and the center of charges in the bottom layer it follows: 
\begin{eqnarray}
\Dblg=D\left(\frac{1}{3-p}+\frac{1-p}{3}\right)
\end{eqnarray}
This allows to write the gap in the top and bottom bilayer and the band-offset $\Deltabands$ :
\begin{eqnarray}
\Deltat=b\Dt+\Deltazero\\
\Deltab=b\Db-\Deltazero\\
\Deltabands=b\Dblg
\end{eqnarray}
Here, we have introduced $\Deltazero$ to take into account the gap that is present without the application of an external field\cite{Rickhaus2019b,Haddadi2019} as well as the displacement field to gap conversion factor $b=\SI{59}{meV/(V/nm)}$, see Methods.
The energy range of the double band region is given by $\Edouble=\Deltabands-\Deltat/2-\Deltab/2=\Deltabands-bD/(3-p)$. We can write:
\begin{eqnarray*}
	\Edouble=\pm bD\left(\frac{1-p}{3}\right)=:bD\cdot f(p)
\end{eqnarray*}
Note that, for $p=1$ (full equilibration of the inner two layers), $f(p)=0$, i.e. the bands do not overlap. \\
Now we need to take into account that the band-structure is gapped at zero displacement field. To enter the double band regime, a certain displacement field, $D_0$, needs to be applied such that the bands overlap. Therefore:
\begin{eqnarray*}
	\Edouble^{+t}&=&bf(p)\cdot(D-D_0)	\qquad \mathrm{for}\: D>D_0\\
	\Edouble^{-t}&=&bf(p)\cdot(D+D_0) 	\qquad \mathrm{for}\: D<-D_0
	\label{eq:Edouble}
\end{eqnarray*}
The density of electrons in the double band regime is $\ndouble=2\Edouble\Cq$, with $\Cq=e^2\partial n/\partial E_\mathrm{F}=e^2 2m^*/\hbar^2\pi$ for a parabolic dispersion. Starting from $E=0$, where $n=0$ we find: 
\begin{eqnarray}
\ndouble^{+t}&=&\Cq bf(p)\cdot(D-D_0)	\qquad \mathrm{for}\: D>D_0\\
\ndouble^{+b}&=&-\Cq bf(p)\cdot(D-D_0) 	\qquad \mathrm{for}\: D>D_0\\
\ndouble^{-t}&=&\Cq bf(p)\cdot(D+D_0)	\qquad \mathrm{for}\: D<-D_0\\
\ndouble^{-b}&=&-\Cq bf(p)\cdot(D+D_0) 	\qquad \mathrm{for}\: D<-D_0
\label{eq:ndouble}
\end{eqnarray}
If $\DeltaCF\approx-bf(p)D_0$, as the measurement suggests, then:
\begin{eqnarray}
\ndouble^{+t}&\approx&\Cq( bf(p) D+\DeltaCF)	\qquad \mathrm{for}\: D>D_0\\
\ndouble^{+b}&\approx&-\Cq( bf(p)D+\DeltaCF)	\qquad \mathrm{for}\: D>D_0\\
\ndouble^{-t}&\approx&\Cq (bf(p)\cdot D-\DeltaCF) \qquad \mathrm{for}\: D<-D_0\\
\ndouble^{-b}&\approx&-\Cq( bf(p)\cdot D-\DeltaCF)	\qquad \mathrm{for}\: D<-D_0
\label{eq:ndouble2}
\end{eqnarray}

\begin{figure}[ht]
	\centering
	\includegraphics[width=1\textwidth]{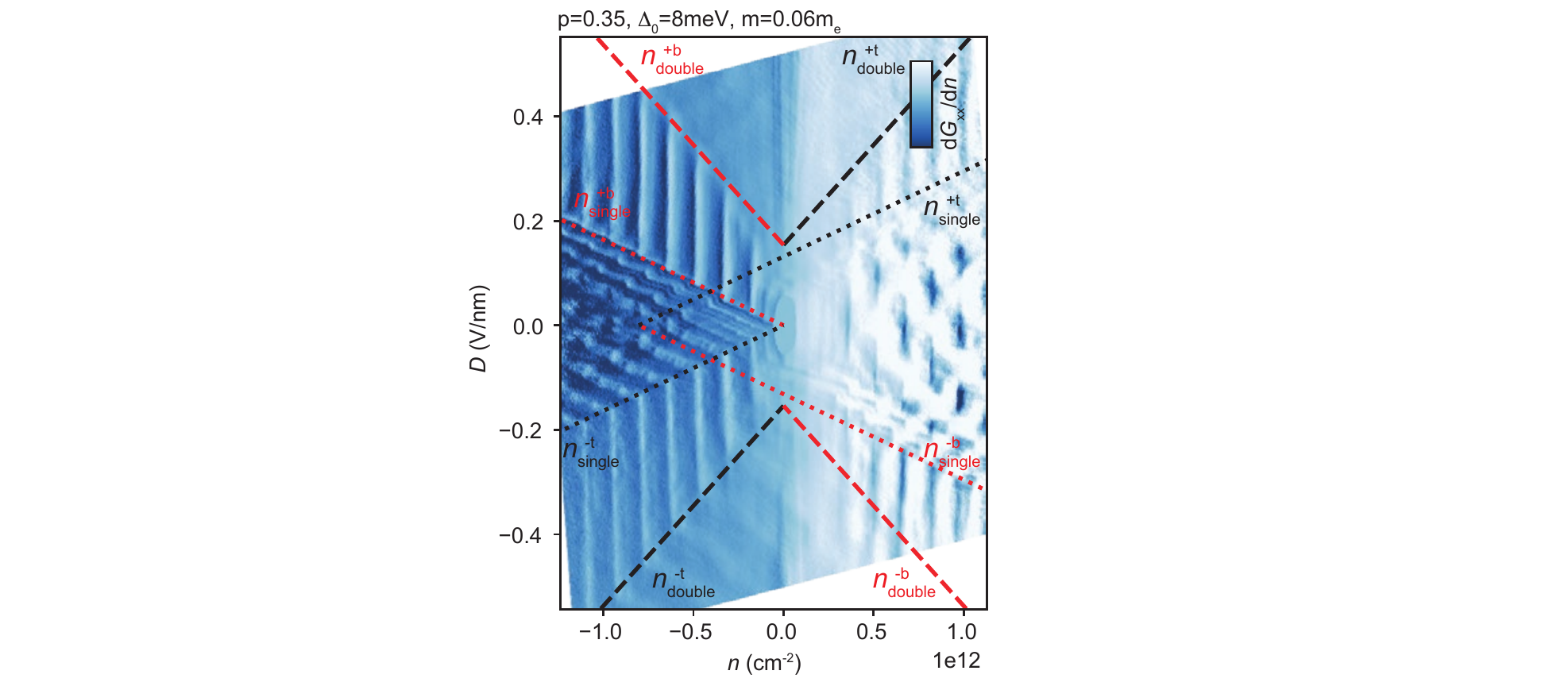}
	\caption{Calculated lines that separate the different regions in the $D$ vs. $n$ map.
	}\label{suppfig:electrostatic-results}
\end{figure}
To reach the end of the single-band regime, $\Esingle$ in the top layer:
\begin{eqnarray*}
	\Esingle^{+t}=\Edouble^+/2+\Deltat
\end{eqnarray*}
In density
\begin{eqnarray*}
	\nsingle^{+t}=\Cq(\Edouble^++\Deltat)  \qquad \mathrm{for}\: D>D_0\\
\end{eqnarray*}
This gives the four lines:
\begin{eqnarray}
\nsingle^{+t}&=&\Cq/e\left( bf(p)\cdot(D-D_0)+b\tilde{f}(p)D+\DeltaCF\right)	\qquad \mathrm{for}\: D>D_0 \\
\nsingle^{-t}&=&\Cq/e\left( bf(p)\cdot(D+D_0) +b\tilde{f}(p)D+\DeltaCF\right)			\qquad \mathrm{for}\: D<-D_0 \\
\nsingle^{+b}&=&-\Cq/e\left(bf(p)\cdot(D-D_0)+b\tilde{f}(p)D-\DeltaCF\right)			\qquad \mathrm{for}\: D>D_0 \\
\nsingle^{-b}&=&-\Cq/e\left( bf(p)\cdot(D+D_0) +b\tilde{f}(p)D-\DeltaCF\right)		\qquad \mathrm{for}\: D<-D_0 \\
\label{eq:nall}
\end{eqnarray}
with:
\begin{eqnarray}
f(p)=\frac{1-p}{3},	\qquad		\tilde{f}(p)=\frac{1}{3-p}, \qquad		g(p)=f+\tilde{f} 
\end{eqnarray}
If $\DeltaCF\approx-bf(p)D_0$, as the measurement suggests, then:
\begin{eqnarray}
\nsingle^{+t}&\approx&\frac{\Cq b g(p)}{e}D-\frac{2\Cq}{e}\DeltaCF 	\qquad\mathrm{for}\: D>D_0 \\
\nsingle^{+b}&\approx&-\frac{\Cq b g(p)}{e}D					\qquad \qquad\qquad \qquad \mathrm{for}\: D>D_0 \\
\nsingle^{-t}&\approx&\frac{\Cq b g(p)}{e}D		 	\qquad \qquad\qquad \qquad\mathrm{for}\: D<-D_0 \\
\nsingle^{-b}&\approx&-\frac{\Cq b g(p)}{e}D-\frac{2\Cq}{e}\DeltaCF	 	\qquad\mathrm{for}\: D<-D_0 \\
\label{eq:nallsimple}
\end{eqnarray}
The results are plotted in Fig.~\ref{suppfig:electrostatic-results}.

\subsection{Zeeman splitting}\label{sec:zeemann}
\begin{figure}[ht]
	\centering
	\includegraphics[width=1\textwidth]{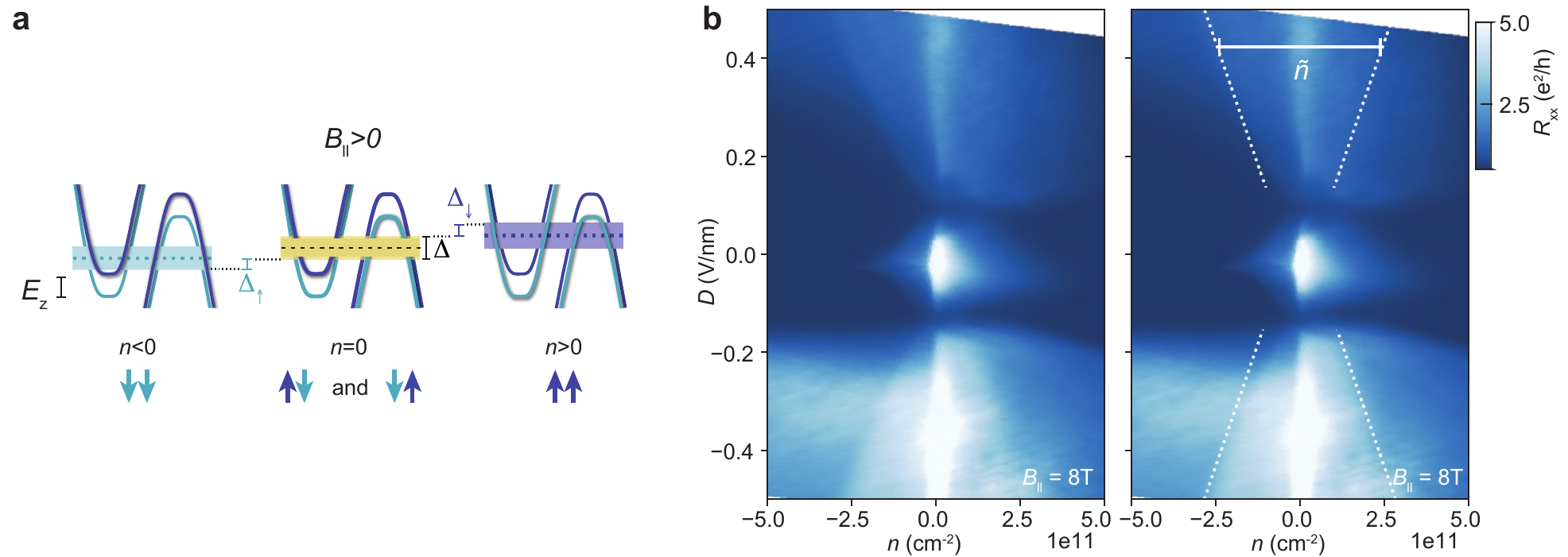}
	\caption{a) Schematics of band structures and gap formation at different $n$ in finite $\Bparallel$ (compare to Fig.~\ref{fig:4}e).
	b) Experimental data without and with fits that determine the expected boundary of the single-spin region (compare to Fig.~\ref{fig:4}g). Dashed lines are coarse fits. The width $\tilde{n}$ of the single-spin region depends on the gap size which scales with $D$.
	}\label{suppfig:Zeemann}
\end{figure}
In the main text we discuss that, at $n=0$, a correlated gap out of opposite magnetic moments can be formed. We now calculate the density $\tilde{n}$ that determines the boundaries of these three gapped regions as a function of $\Bparallel$ and $D$ (see Fig.~\ref{suppfig:Zeemann}b).

By applying a parallel magnetic field $\Bparallel$, The energy of bands with different magnetic moments shift by the Zeeman energy $\Ez=g\muB\Bparallel$, with $g=2$ in graphene.
We label the regions where only one spin band is gapped as $\Deltaup$ and $\Deltadown$ (see Fig.~\ref{suppfig:Zeemann}a). The size of these regions is given by the Zeemann energy $\Deltaup+\Deltadown=\Ez$. The energy range of the band structure that is partially or entirely gapped is therefore $\Ez+\Delta$, where $\Delta$ corresponds to the gap size at $n=0$. We thus obtain $\tilde{n}$:
\begin{eqnarray}
    \tilde{n}= \frac{\Cq}{e^2}(2\Delta + \Ez)
\end{eqnarray}
where $\Cq=e^2\cdot2\meff/(\hbar^2\pi)$ for a parabolic bilayer band. When converting gap to density, $\Delta$ has to be multiplied by 2 since all the bands are gapped there, as opposed to the single-spin gapped regions (see Fig.~\ref{suppfig:Zeemann}). 

To obtain the coarse fits we estimate $\Delta(D,\Bparallel)$ from the experiment. In Fig.~\ref{fig:4}b we can roughly fit the size of the correlated gap:
\begin{eqnarray}
\Delta(D,\Bparallel=0)=\alpha(D-D_0)
\end{eqnarray}
with $\alpha=\SI{15}{meV/(Vnm^{-1})}$ and $D_0=\SI{0.15}{V/nm}$. From bias measurements at $n=0$ as a function of $\Bparallel$ (Fig.~\ref{suppfig-B=}) we find that $\Delta(\Bparallel=\SI{8}{T})/\Delta(\Bparallel=0)\approx1/2$. We thus obtain $\tilde{n}(D,\Bparallel=\SI{8}{T})$. With $\meff=0.12\me$ we obtain the dashed line shown in Fig.~\ref{suppfig:Zeemann}b.

\subsection{Gap in parallel magnetic field}\label{sec:gapinBparallel}
\begin{figure}[ht]
	\centering
	\includegraphics[width=1\textwidth]{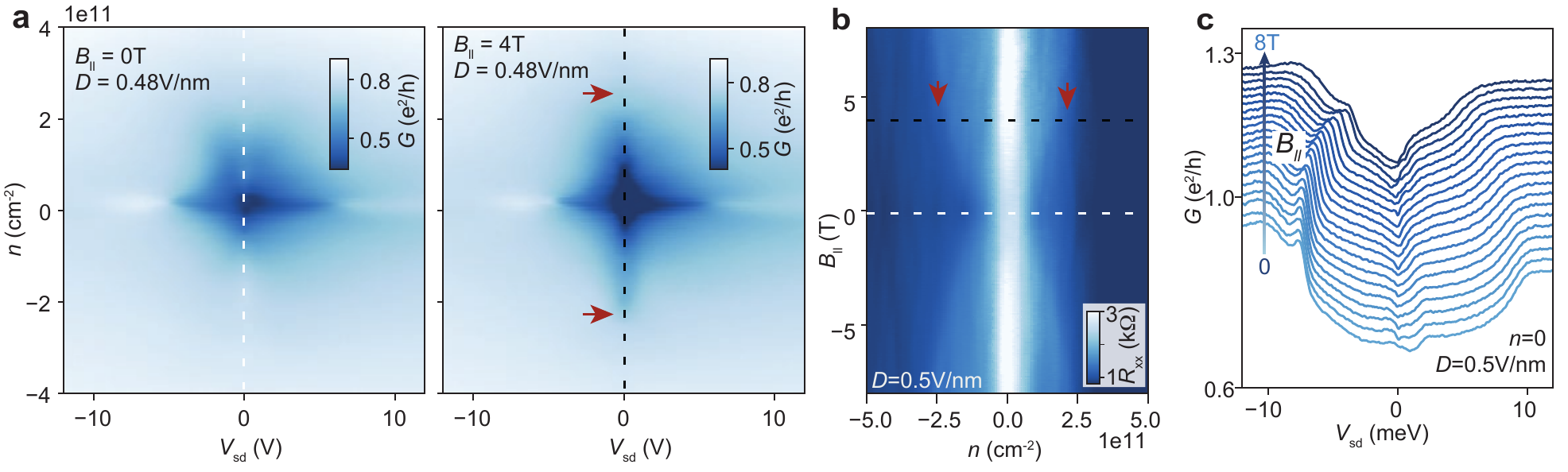}
	\caption{a) $G(\Vsd,n)$ at large $D$ for $\Bparallel=0$ and $\Bparallel=\SI{4}{T}$. With red arrows we mark the end of the single-band region that occurs due to Zeeman splitting.
		b) As a comparison, we mark this region also in $\Rxx(n,\Bparallel)$, as well as the corresponding cuts in a)
		c) Gap in $G(\Vsd)$ as a function of $\Bparallel$.
	}\label{suppfig-B=}
\end{figure}
Here we show additional data in parallel magnetic field. In Fig.~\ref{suppfig-B=}a we show the gap in a $\Vsd$ measurement as a function of density and for $\Bparallel=0$ and $\Bparallel=\SI{4}{T}$. As a guide, we mark the corresponding lines in the $\Rxx$ map (Fig.~\ref{suppfig-B=} and Fig.~\ref{fig:4}f). The source drain measurement exhibits a reduced conductance at  $\Bparallel=\SI{4}{T}$ between the region, marked by red arrows, which is not present at $\Bparallel=0$. We attribute this to a partially gapped band structure, as argued previously. 

In Fig.~\ref{suppfig-B=}c, we show the evolution of the gap with $\Bparallel$ at $\Dmax$ and $n=0$ in a $\Vsd$ waterfall plot. With increasing $\Bparallel$, the feature that we identify as the main gap is reduced in energy. This is consistent with the Zeeman-split spin bands (Fig.~\ref{suppfig:Zeemann}a), where two gaps of different size emerge for the two possibilities of spin pairing at $n=0$. Note that in the bias measurement, the region where all spin bands are gapped (given by the size of the smallest gap) leads to the most significant decrease of conductance. I.e. the gap that is observed in Fig.~\ref{suppfig-B=}c corresponds to the smaller gap at $n=0$, which is decreasing with $\Bparallel$. 

\subsection{Low-energy continuum model of TDBG}\label{sec:model_H}
We construct the low-energy continuum model of TDBG by generalizing the well-established Bistritzer-MacDonald model\cite{MATBG} to TDBG case. We consider the ABAB-stacked TDBG which is relevant to this specific experiment and assume that the top bilayer (layer 1 and 2) is twisted anticlockwise by $\theta/2$ and the bottom bilayer (layer 3 and 4) is twisted clockwise by $\theta/2$. 
The valley- and spin-projected Hamiltonian acting on eight-component (four layers and two sublattices) $\pmb{k} \cdot \pmb{p}$ spinors $\Psi = (a_{\text{1A}}, a_{\text{1B}}, a_{\text{2A}}, a_{\text{2B}}, a_{\text{3A}}, a_{\text{3B}}, a_{\text{4A}}, a_{\text{4B}})^{\text{T}}$ is 
\begin{equation} \label{Eq_Hamiltonian}
H(\pmb{k}) = 
\begin{pmatrix}
    h^{(1)}_{\theta/2}(\pmb{k}) & T_{\text{Bernal}} & 0 & 0 \\
    T_{\text{Bernal}}^\dagger & h^{(2)}_{\theta/2}(\pmb{k}) & T & 0 \\
    0 & T^\dagger & h^{(3)}_{-\theta/2}(\pmb{k}) & T_{\text{Bernal}} \\
    0 & 0 & T_{\text{Bernal}}^\dagger & h^{(4)}_{-\theta/2}(\pmb{k})
\end{pmatrix}
\end{equation}
$h_{\theta}(\pmb{k})$ is the Dirac Hamiltonian rotated by $\theta$,
\begin{equation}
    h^{(l)}_\theta(\pmb{k}) = 
    \begin{pmatrix}
        \varepsilon_{l\text{A}} & \xi \hbar v_\text{F} q e^{-i\xi(\theta_{\pmb{q}}-\theta)} \\
        \xi \hbar v_\text{F} q e^{i\xi (\theta_{\pmb{q}}-\theta)} & \varepsilon_{l\text{B}}
    \end{pmatrix}
\end{equation}
where $\varepsilon_{l\text{A}}$ and $\varepsilon_{l\text{B}}$ are on-site energies of layer $l$, $\xi=\pm 1$ represent two valleys, $\pmb{q} = \pmb{k} - \xi \pmb{K}$ is the momentum measured from graphene Brillouin-zone corner and the Fermi velocity $v_\text{F}=10^6$ m/s is adopted in our calculations throughout the paper. The tunneling between Bernal-stacked bilayer is
\begin{equation}\label{Eq_TBernal}
T_{\text{Bernal}} = 
    \begin{pmatrix}
    0 & \gamma_1 \\
    0 & 0
    \end{pmatrix}
\end{equation}
where only the tunneling, $\gamma_1=330$ meV, between the dimer sites is retained. Non-dimer sites tunnelings are not important for low-energy bands of intermediate twist angles. The interlayer tunneling between the middle two graphene layers is the same as in Bistritzer-MacDonald model\cite{MATBG},
\begin{equation}
    T_{\pmb{k}', \pmb{k}} = w \sum\limits_{j=1}^3 \delta_{\pmb{k}', \pmb{k}-\pmb{q}_j} T_j
\end{equation}
To account for the corrugation and strain effects, the interlalyer tunneling between the same sublattice is reduced\cite{Yoo_corrugation}, $w_{\text{AA}} = 0.8 w = 0.8 w_{\text{AB}} = 88$ meV.

\subsection{Asymmetry of the single-band regions}\label{sec:asymmetry}
\begin{figure}[ht]
	\centering
	\includegraphics[width=1\textwidth]{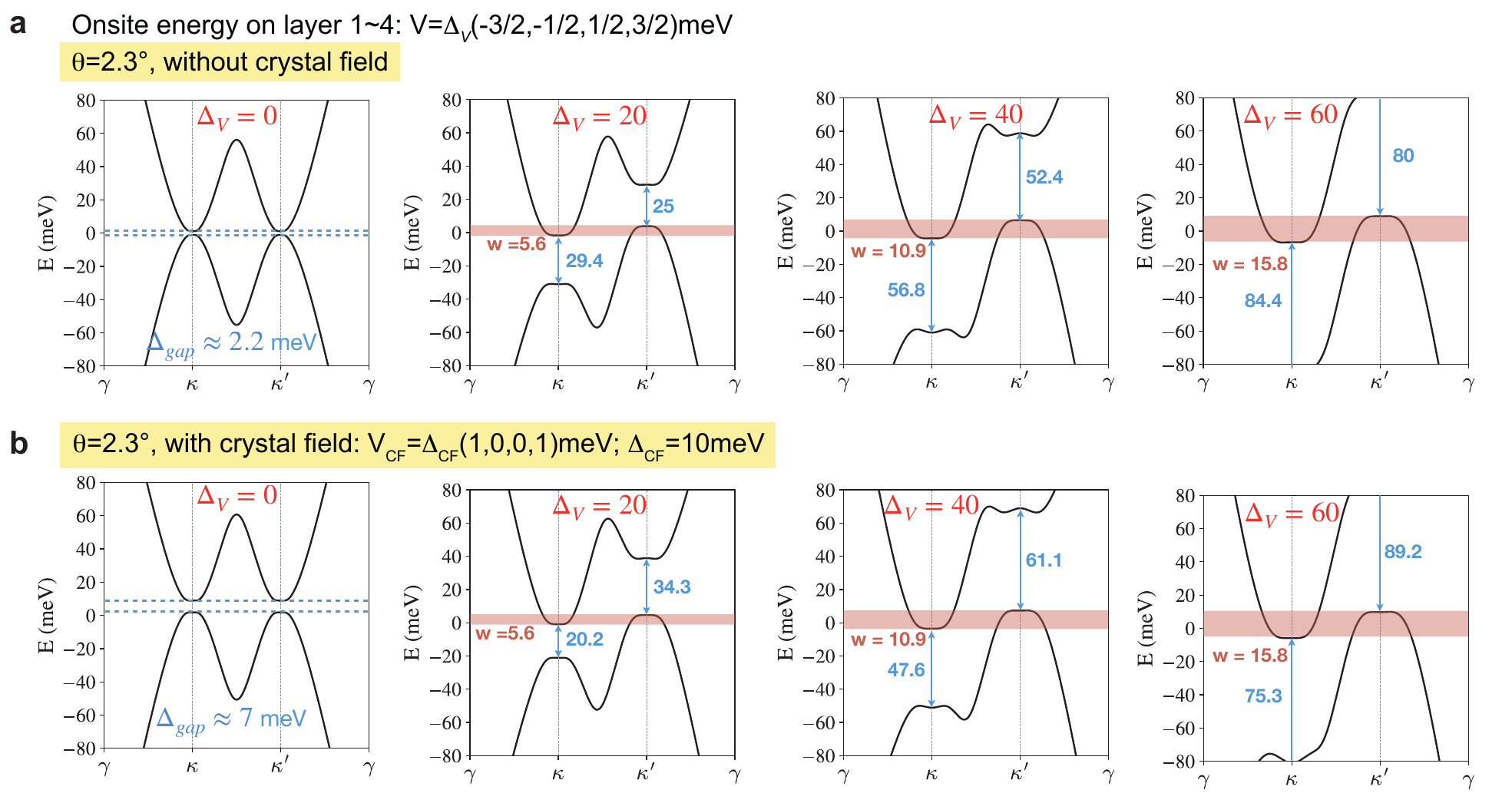}
	\caption{Valley-projected band structures of $2.3^\circ$-TDBG for different displacement fields $D\propto\Delta_V$ using the low-energy continuum model. (a) Without crystal field contribution. (b) With the crystal field contribution.
	}\label{suppfig:asymmetry}
\end{figure}
Consider a cut in Fig.~\ref{fig:2}e of the main text for $D=\SI{0.3}{V/nm}$. In the experiment, for $D>0$, The single band-region of the valence band in the top bilayer (light blue) spans a larger density range than the single-band region of the conduction band in the bottom bilayer (light purple). 
However, in the calculation for $\Delta_V>0$ in Fig.~\ref{fig:2}gh and Fig.~\ref{suppfig:asymmetry}b, the asymmetry is opposite, i.e. the gap at $\kappa$ (light blue) spans a smaller energy range than the gap at $\kappa'$ (light purple).

The origin of this asymmetry is closely linked to the gap that can be observed without external field ($D=0$ or $\Delta_V=0$ in Fig.~\ref{suppfig:asymmetry}). At larger twist angles, this gap is solely due to a spontaneous layer polarization of charge carriers, since the inner layers have a different electrostatic environment from the outer layers. This layer polarization has been measured and theoretically confirmed \cite{Rickhaus2019b} at larger twist angles, but also needs to be taken into account here. In Fig.~\ref{suppfig:asymmetry}a and b we compare band structures without and with the crystal field contribution. Apparently, the asymmetry changes, when crystal fields are taken into account. The gap for positive energies is larger than the one for negative ones if crystal fields are considered, in contradiction with the experimental findings at $\twist$ but in agreement with the large twist angle device, Fig.~\ref{suppfig:twists}a and ref. \cite{Rickhaus2019b}.

It is to be noted, however, that the band structure is also gapped if the crystal field contribution is not taken into account, see Fig.~\ref{suppfig:asymmetry}. Importantly, upon changing the interlayer bias $\Delta_V$, this leads to an opposite asymmetry, counteracting the crystal field effect. We argue, that the continuum model may underestimate this effect. In the following, we present indications that in the experiment, the spontaneous layer polarization of charges at $D=0$ by the moiré lattice is stronger than captured by the continuum model. 

There are other effects that the model does not capture very accurately, e.g. the density at which the van-Hove singularity (VHS, $\sim\SI{60}{meV}$ in Fig.~\ref{suppfig:asymmetry}a) is to be expected in the conduction and valence band. Whereas the bands for $\Delta_V=0$ are almost perfectly electron-hole symmetric in the calculation (both with and without crystal fields), we observe a significant difference in densities of the VHS in the conduction ($n=\SI{5.8e12}{cm-2}$) and valence band ($n=\SI{5e12}{cm-2}$). 

This qualitatively agrees with the larger effective mass that we observed for the large angle twisted bilayer device \cite{Rickhaus2019b}, where the ratio of effective masses in the valence/conduction band was $m_v/m_c\sim1.5$ in the experiment (though with a significant error bar), $m_v/m_c\sim1.2$ in the DFT calculation and $m_v/m_c\sim1.1$ in the tight binding calculation \cite{Rickhaus2019b}. We conclude that modeling the electron-hole asymmetry accurately on a quantitative level appears to be difficult using single-particle band structure calculations. However, the strength of the asymmetry of the gaps upon application of an interlayer bias will strongly depend on the electron-hole symmetry.

Finally, our tight binding model does not include lattice relaxation effects, which can have an impact on the asymmetry and the crystal field gap \cite{Haddadi2019}. Since the energy scales of the asymmetry are rather small, a quantitative agreement between experiment and tight binding calculations is not to be expected, but not required either.

\subsection{Hartree-Fock calculations}

\begin{figure}
    \centering
    \includegraphics[width=0.98\textwidth]{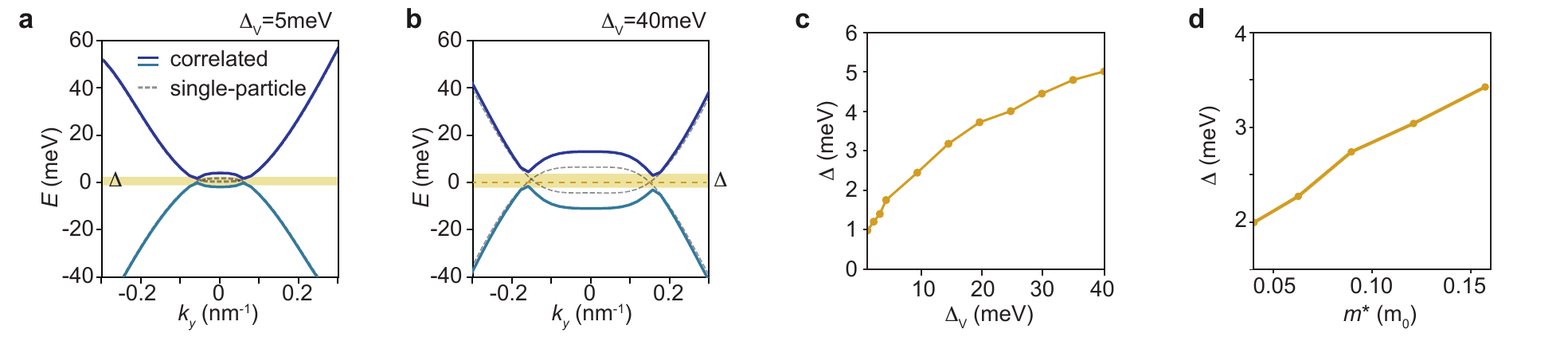}
    	\caption{\textbf{Gap opening in HF calculations}. 
    	a) Single-particle (dashed) and HF  (solid) bands at $\DeltaV=\SI{5}{meV}$ and
        b) $\DeltaV=\SI{40}{meV}$ at $\twisttheory$.
        c) The correlated gap $\Delta$, obtained from HF calculations, induced by intervalley Fermi surfaces nesting as a function of $\Delta_V$. We reproduce the experimental observations. The gap induced by intravalley Fermi surfaces nesting is shown in Fig.~\ref{suppfig:valley}b for comparison.
        d) Correlated gap $\Delta$ strongly depends on $\meff$. $m^*$ is tuned by adjusting the parameter $\gamma_1$ in the Bernal-stacked bilayer graphene Hamiltonian Eq.(\ref{Eq_TBernal}) in SI \ref{sec:model_H}. This figure is calculated with $\theta=2.3^\circ$, $w=0$, $\Delta_V=15$ meV and $v_\text{F}=10^6$ m/s. 
    }\label{fig:5}
    \end{figure}

The results of the Hartree-Fock calculations that are discussed in the main text are shown in Fig.~\ref{fig:5}. A gap opening between the energetically overlapping electron and hole bands  due to correlations is confirmed.

\subsection{Fermi surfaces and nesting momenta}\label{sec:valley}
\begin{figure}[ht]
	\centering
	\includegraphics[width=1\textwidth]{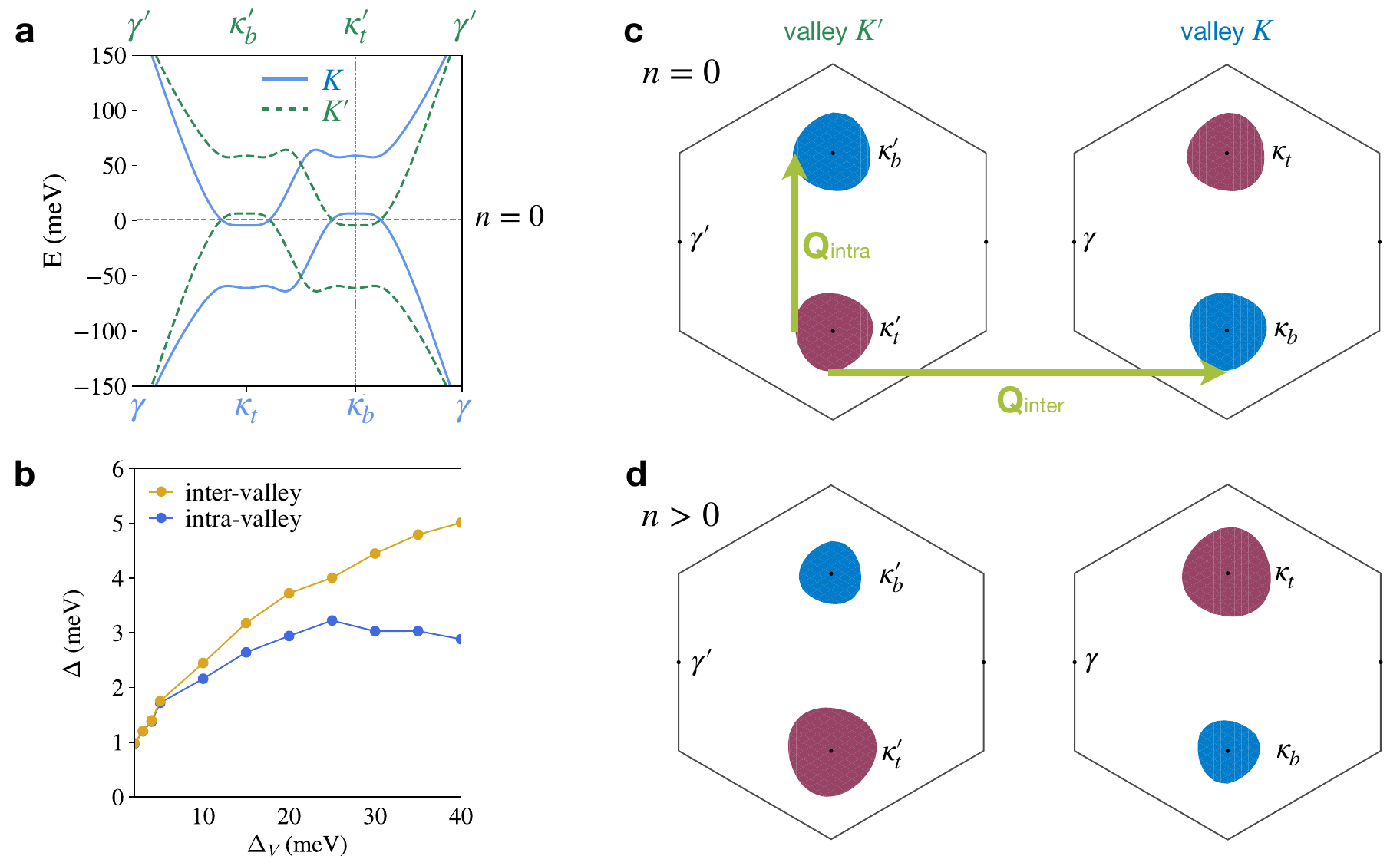}
	\caption{
	a) Dispersion relations of $2.3^\circ$-TDBG for $\DeltaV=\SI{40}{meV}$ in the $K$ (solid) and $K'$ (dashed) valleys. Crystal fields are absent.
	b) Correlated gap $\Delta$ {\it vs.} $\Delta_V$ for intervalley nesting (yellow) and intravalley nesting (blue).
	c) Fermi surfaces at $E=0$, $n=0$ and 
	d) $E>0$, $n>0$. The green arrows show an intra- ($\textbf{Q}_\mathrm{intra}$) and an inter-valley ($\textbf{Q}_\mathrm{inter}$) nesting wavevector.
	}\label{suppfig:valley}
\end{figure}
In Fig.~\ref{suppfig:valley}a we show the dispersion relation at $\DeltaV=40$ meV in the two valleys $K$ and $K'$. We label the moir\'e Brillouine zone corners of the top and bottom bilayers in valley $K$ as $\kappa_t$ and $\kappa_b$ respectively, and label in valley $K'$ as $\kappa_t'$ and $\kappa_b'$. The corresponding electron (red) and hole (blue) Fermi-surfaces are shown in Fig.~\ref{suppfig:valley}c at charge neutrality n=0.
With green arrows we depict two wavevectors, one corresponding to intra-valley nesting ($\textbf{Q}_\mathrm{intra}$) and another corresponding to inter-valley ($\textbf{Q}_\mathrm{inter}$) nesting. Since the Fermi surfaces around the $\kappa_t$ and $\kappa_b$ in the same valley are not perfectly circular, it is not possible to connect the two Fermi surfaces in the same valley with the same $\textbf{Q}_\mathrm{intra}$. It is, however, possible to connect the Fermi surfaces between different valleys with $\textbf{Q}_\mathrm{inter}$. This suggests that the correlated state is most stable when it pairs electrons and holes in opposite valleys,  

{\it i.e. }$\kappa_b$ $\xleftrightarrow{}$
$\kappa_t'$ and $\kappa_t$ $\xleftrightarrow{}$
$\kappa_b'$, and
\begin{equation}
\begin{split}
    \kappa_{t/b} &= \mathcal{R}_{t/b} \pmb{K}_j  + n\pmb{b}_1 + m\pmb{b}_2 \\
    \kappa'_{t/b} &= \mathcal{R}_{t/b} \pmb{K}'_j + n'\pmb{b}_1 + m'\pmb{b}_2
\end{split}
\end{equation}
where rotational operators $\mathcal{R}_t = \mathcal{R}(\theta/2)$ and $\mathcal{R}_b = \mathcal{R}(-\theta/2)$. $\pmb{K}_j$($\pmb{K}'_j$) for $j = 1,2,3$ are three graphene's first Brillouin zone corners in valley $K$($K'$). $n,m,n',m'$ are integers. $\pmb{b}_1$ and $\pmb{b}_2$ are moir\'e reciprocal lattice vectors. The density-wave momentum $\pmb{Q}_{\text{inter}}$ is therefore
\begin{equation}
    \pmb{Q}_{\text{inter}} = \pmb{\kappa}'_t-\pmb{\kappa}_b = \mathcal{R}_t \pmb{K}'_{j'} - \mathcal{R}_b \pmb{K}_j + n\pmb{b}_1 + m\pmb{b}_2
\end{equation}
For $n=0$ and $m=0$, there are nine possible $\pmb{Q}_{\text{inter}}$'s: $\pmb{Q}^1_{\text{inter}}$, $\pmb{Q}^2_{\text{inter}}$, $\pmb{Q}^3_{\text{inter}}$ and their $C_3$-rotated vectors, which are schematically shown in Fig.~\ref{fig:nestingvectors} below. 

\begin{figure}[h]
\centering
\includegraphics[scale=0.7]{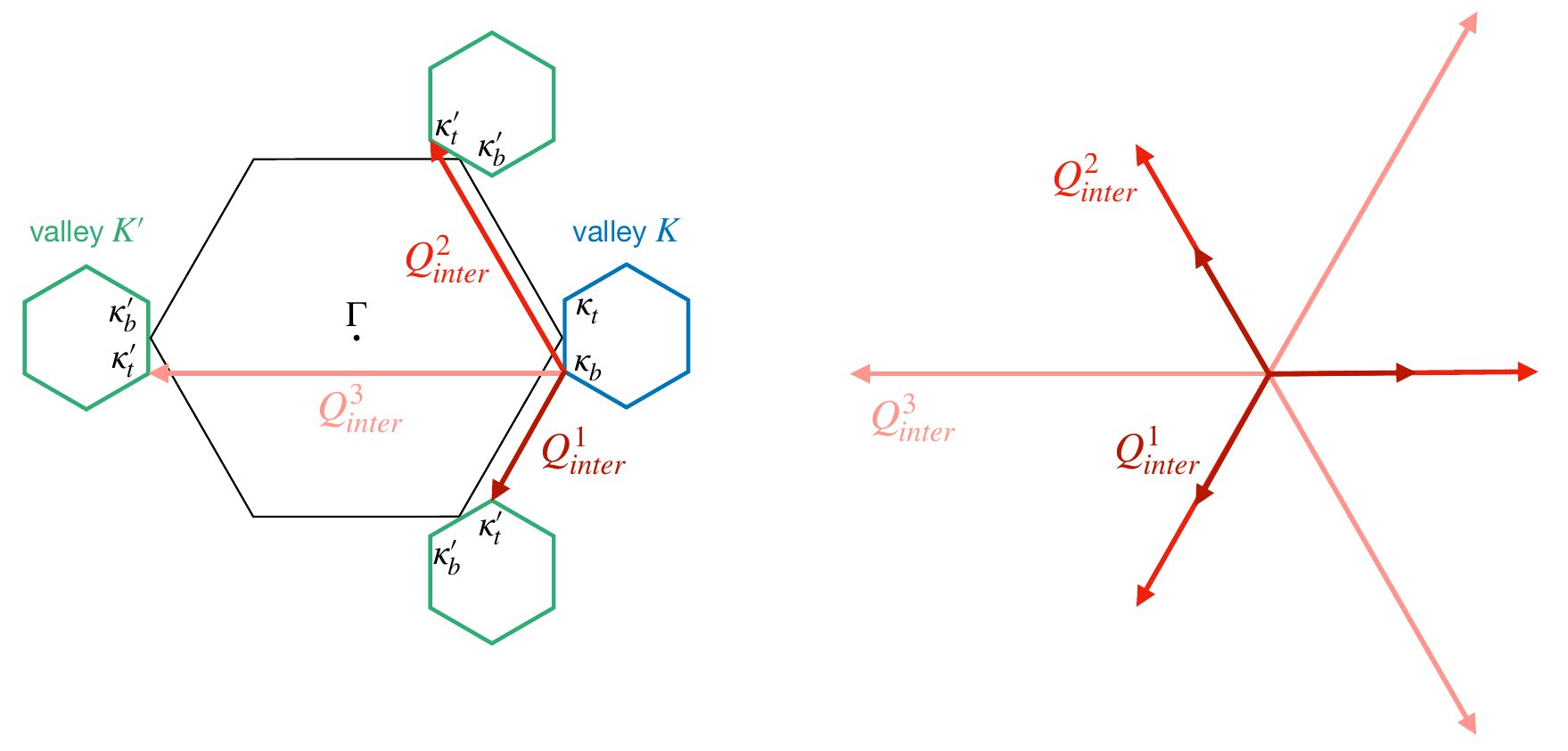}
     \caption{\label{fig:nestingvectors}
     Possible Fermi surface nesting momenta $\pmb{Q}_{\text{inter}}$. Intervalley nestings are favored.}
\end{figure}

At finite densities (Fig.~\ref{suppfig:valley}d), not only the shape but also the size of the Fermi surfaces within the same valley and between opposite valleys become different, explaining the observed vanishing correlated state.

We show the correlated gap $\Delta$ {\it vs.} $\Delta_V$ for intervalley and intravalley nestings in Fig.~\ref{suppfig:valley}(b). At small $\Delta_V$, intervalley- and intravalley-induced gap have the same size as a result of small Fermi surface anisotropy. While at large $\Delta_V$, the intervalley nesting is prefered.

\subsection{DW states in a perpendicular magnetic field}\label{sec:LLs}
\begin{figure}[ht]
	\centering
	\includegraphics[width=1\textwidth]{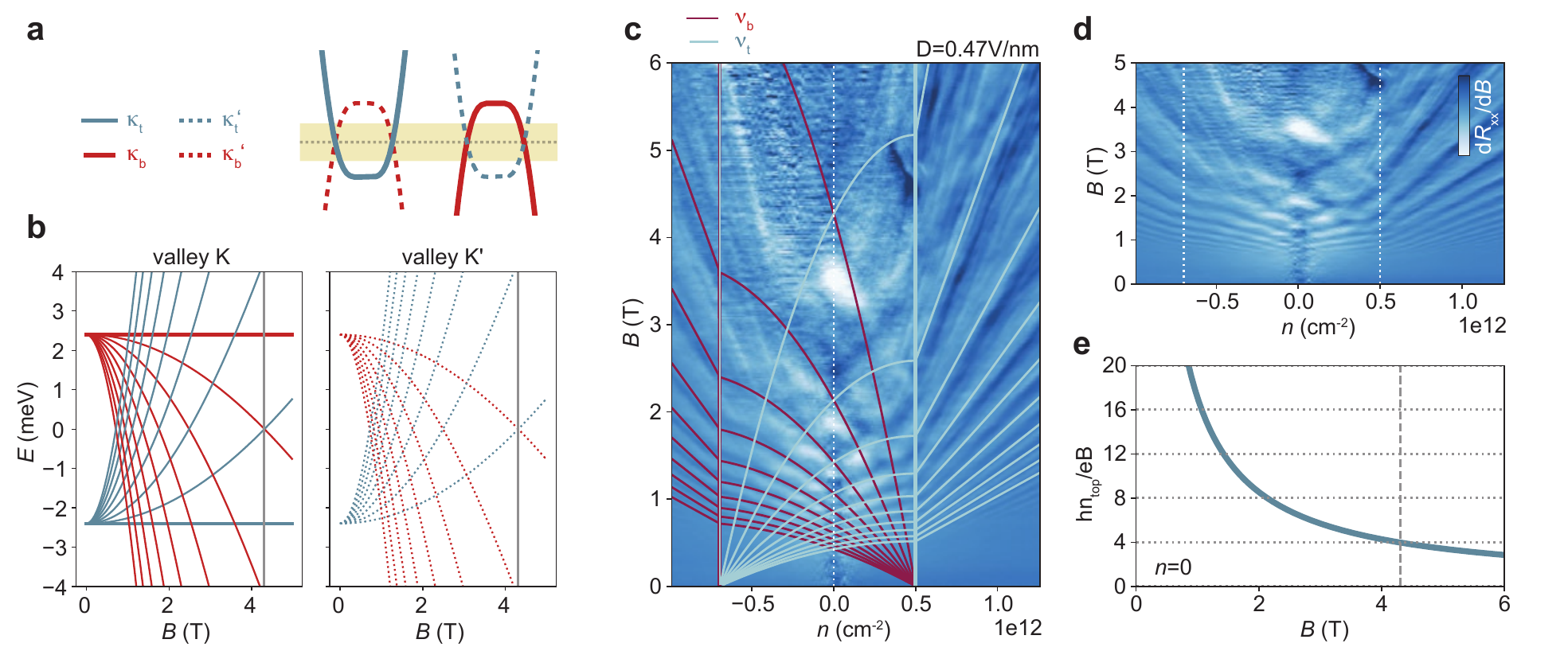}
	\caption{a) Schematic band structures in moiré valleys $\kappa_t$ and $\kappa_b$ (red and turquoise) and microscopic valleys $K$ and $K'$ (solid and dashed).
	b) Landau level dispersions for a given top/bottom bilayer inter bilayer energy offset energy 
	and individual bilayer gap, $\Delta_b$. 
	c)d) Empirically fitted filling factors in the top (turqouise) and bottom (red) bilayer for the data in the main text, Fig.\ref{fig:1}d, but over a larger magnetic field range.
	e) Filling factor in the top layer as a function of $B$.}
	\label{suppfig-LLs}
\end{figure}

The valley g-factors in bilayer graphene have opposite signs for opposite microscopic valleys $K$ and $K'$ (solid and dashed lines in Fig.\ref{suppfig-LLs}a) and opposite signs for electrons and holes.  A perpendicular magnetic field therefore shifts conduction  and valence band 
energies in the same valley in the opposite direction. For positive displacement fields, conduction bands form at $\Kt$ and $\Ktprime$ and valence bands form at $\Kb$ and $\Kbprime$, see Fig.\ref{suppfig-LLs}a. A finite magnetic field shifts $\Kt$ and $\Kbprime$ up and shifts $\Ktprime$ and $\Kb$ down, favoring intravalley nesting over intervalley nesting at $n=0$ condition. The magnetic field is therefore pair-breaking for opposite valley pairing.

The quantum physics behind the valley g-factors is that the anomalous $N=0$ and 1 Landau levels (LLs) are immediately shifted in energy - either to the conduction or valence band side of the gap depending on valley, for arbitrarily weak magnetic fields.
For a bilayer with gap $\Delta_b$ the LL energies are
\begin{eqnarray}
E_N = \pm \sqrt{{\frac{1}{4}} \Delta_b^2+N(N-1) (\hbar \omega_c)^2} &  N = 2,3, \ldots \\
E_N = \frac{1}{2}\tau_z \Delta_b & N=0,1
\end{eqnarray}
where $\tau_z$ equals $\pm 1$ for microscopic valleys $K, K'$. For the dispersion given in Fig.\ref{suppfig-LLs}a (i.e. overlapping electron-hole bands), the evolution of LLs as a function of $\Bperp$ is plotted. Note that the difference between valley $K$ and $K'$ is the presence/absence of the anomalous $N=0,1$
Landau levels whose energies do not depend on field. When $\hbar \omega_c$ is small compared to $\Delta_b$ there is a large gap between $-\Delta_b$ and $\Delta_b$. Including spin, the density in this gap is $-2/(2\pi\ell^2)$ for $\tau_z=1$ and $+2/(2 \pi \ell^2)$ for $\tau_z=-1$, where $\ell=\sqrt{\hbar/eB}$ is the magnetic length. It follows that at $n=0$, the electron and hole energies will be degenerate not for opposite valley pairing, but for like valley pairing. 
The decrease in gap with increasing $\Bperp$ observed in Fig.\ref{fig:5}d) corresponds to transport in the low-carrier density valley.

To understand the origin of the critical field ($B_c=\SI{4}{T}$ at $D=\SI{0.5}{V/nm}$ in Fig.\ref{fig:5}d), the following numbers can be considered. At $\Bperp=\SI{4}{T}$, the extra density from the anomalous Landau level is $\approx\SI{2e11}{cm^{-2}}$. The origins of the Landau fans in Fig.\ref{suppfig-LLs} are at densities $\SI{-7e11}{cm^{-2}}$ and $\SI{5e11}{cm^{-2}}$, therefore the density per layer at $n=0$ is $\approx\SI{3e11}{cm^{-2}}$ and one valley is completely depopulated at $B\approx\SI{6}{T}$. The actual value of magnetic field at which one valley is depopulated can be smaller, however, since the valley splitting is likely enhanced by interactions. Indeed, by empirically fitting the LL transition in Fig.\ref{suppfig-LLs}c we can estimate that filling-factor $\nu=4$ in one of the layers (corresponding to depopulation of one valley) is reached around $\SI{4}{T}$, see Fig.\ref{suppfig-LLs}e.

\subsection{Layer polarization and the effective mass}\label{sec:theorymeff}

\begin{figure}
\includegraphics[width=1\columnwidth]{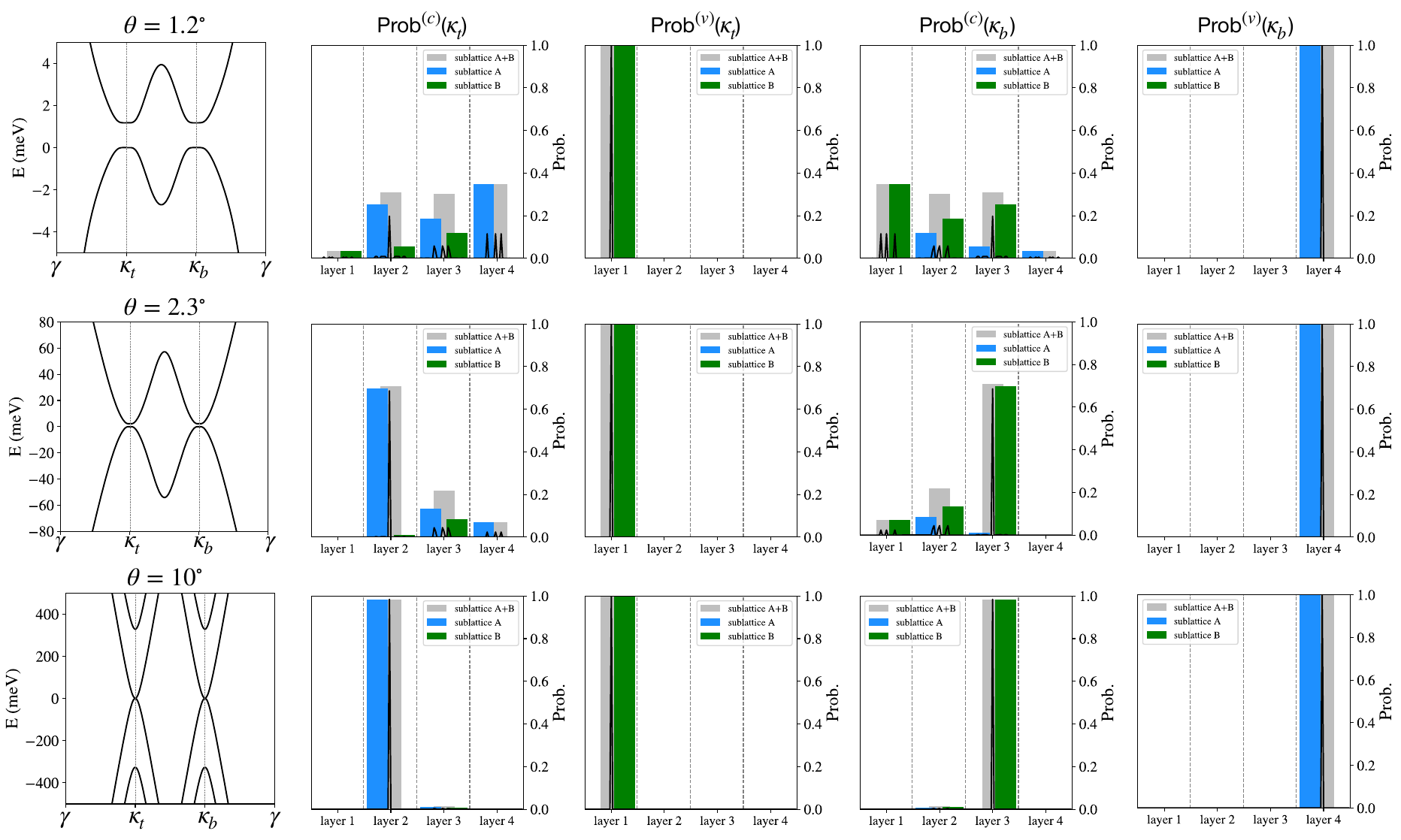}
\centering
\caption{Band structures and wave function probabilities defined in Eq.~(\ref{Eq_prob}) at $\pmb{\kappa}_t$ and $\pmb{\kappa}_b$ of the lowest conduction and highest valence bands. The black curves in probability figures plot $|z_{l,\alpha,\pmb{G}}|^2$.
(a) $\theta=1.2^\circ$, (b) $\theta=2.3^\circ$, (c) $\theta=10^\circ$. For small twist angles, the middle two graphene layers are strongly coupled thereby decreasing the layer polarization. For $\theta \gtrsim 2^\circ$, the wave function is mainly localized on a specific sublattice and a moir\'e reciprocal lattice vector $\mathbf{G}_0=0$ (indicated by the sharp black peaks in probability figures).}
\label{suppfig:probk}
\end{figure}

We show single-particle band structures and probabilities to find a band state on each layer at moir\'e Brillouin zone corners $\pmb{\kappa}_t$ and $\pmb{\kappa}_b$ in valley $K$ for twist angles $1.2^\circ$, $2.3^\circ$ and $10^\circ$ in Fig.~\ref{suppfig:probk}, where crystal fields and external electric fields are ignored. The probability on layer $l$ and sublattice $\alpha$ is defined to be 
\begin{equation}\label{Eq_prob}
\text{Prob}^{(n)}_{l,\alpha}(\pmb{k}) = \sum\limits_{\pmb{G}}|z^{(n)}_{l,\alpha,\pmb{G}}(\pmb{k})|^2
\end{equation}
where $n$ is a band index, $\pmb{G}$ are moir\'e reciprocal lattice vectors and $z$ is the wavefunction of Hamiltonian in Eq.~(\ref{Eq_Hamiltonian}).

As the twist angle increases, electrons are concentrated on layer 2 at $\pmb{\kappa}_t$ and on layer 3 at $\pmb{\kappa}_b$. Here layer 1 to 4 denote top-most layer to bottom-most layer as shown in Fig.~\ref{fig:1}a. Since the probabilities of valence bands are invariant with respect to the twist angle, which is a property of Eq.~(\ref{Eq_TBernal}) ignoring the tunneling between non-dimer sites, we define the layer polarization as the probability on the top bilayer (layer 1,2) at $\pmb{\kappa}_t$, {\it i.e.} $\probk$:

\begin{equation}
    \probk = \sum\limits_{\pmb{G},l=1,2,\alpha=\text{A,B}}|z^{(\text{c})}_{l,\alpha,\pmb{G}}(\pmb{\kappa}_t)|^2
\end{equation}

where the superscript c denotes the lowest conduction band.

To qualitatively capture the band flatness near the Fermi level, we estimate the effective mass $m^*$ by 
\begin{equation}\label{Eq_effmass}
m^* = \frac{\hbar^2(\pmb{k}-\pmb{\kappa})^2}{2m_e(E_{\pmb{k}}-E_{\pmb{\kappa}})}
\end{equation}
and $\pmb{k}$ is taken to be $|\pmb{\kappa}_b-\pmb{\kappa}_t|/4$ away from the reference point $\pmb{\kappa}$.

\begin{figure}
\includegraphics[width=1\columnwidth]{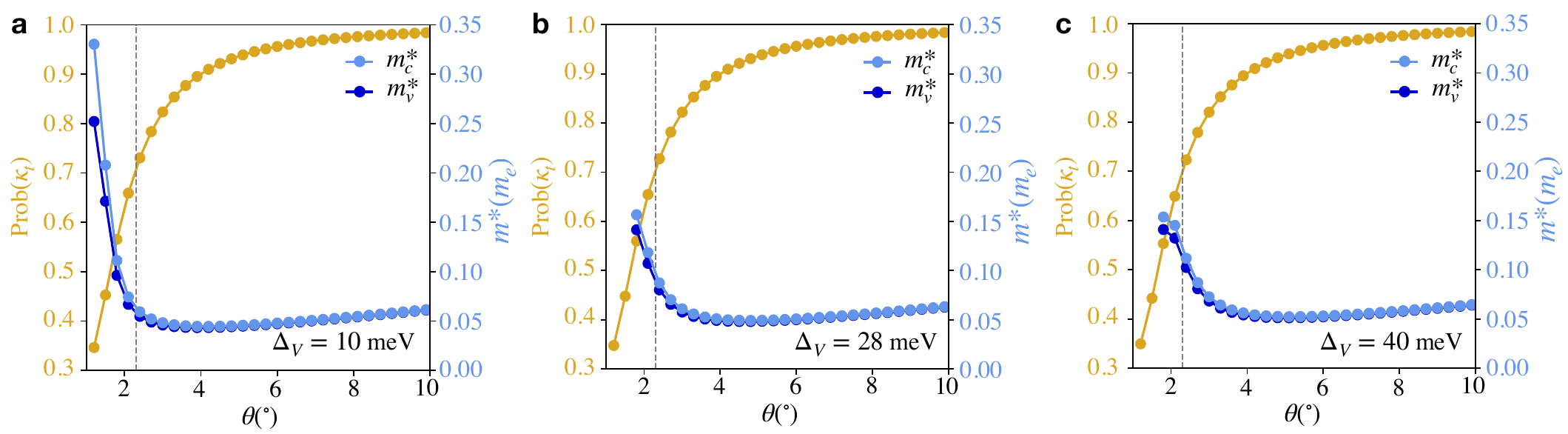}
\centering
\caption{Layer polarization Prob($\kappa_t$) and effective mass $m^*$ in units of free electron mass $m_e$ with respect to twist angle. The crystal fields are absent. $m^*_c$ ($m^*_v$) is the effective mass of conduction (valence) band near $\kappa_t$ ($\kappa_b$) in valley $K$. Some data points of $m^*$ at small $\theta$ are absent, that is because at small twist angles and under relatively large displacement field ($\Delta_V$), the low-energy bands are too flat to use the approximation in Eq.~(\ref{Eq_effmass}). The vertical dashed line is a guide to the eye at $\theta=2.3^\circ$, which is the relevant angle in our experiment and theoretical calculations. (a) $\Delta_{\text{V}}=10$ meV, (b) $\Delta_{\text{V}}=28$ meV, (c) $\Delta_{\text{V}}=40$ meV. At some finite $\Delta_{\text{V}}$, both layer polarization and effective mass are large near $\theta \sim 2^\circ$. This explains our observation of the excitonic insulating state only near an intermediate twist angle $\sim2^\circ$.}
\label{suppfig:prob_mass}
\end{figure}

Figure~\ref{suppfig:prob_mass} shows the layer polarization Prob($\kappa_t$) and effective masses of conduction band $m^*_c$ and valence band $m^*_v$ as a function of twist angle for external fields $\Delta_{\text{V}}=10$, $\Delta_{\text{V}}=28$ and $\Delta_{\text{V}}=40$ meV, ignoring the crystal fields. $m^*_c$ ($m^*_v$) is estimated near $\kappa_t$ ($\kappa_b$) as defined in Eq.(\ref{Eq_effmass}). The layer polarization increases quickly with twist angle and does not evidently depend on  $\Delta_\text{V}$. On the contrary, the effective masses decrease quickly with twist angle and the steepness depends on $\Delta_{\text{V}}$. At some finite $\Delta_{\text{V}}$, the situation is optimized near $\theta \sim 2^\circ$ where both layer polarization and effective mass are large. This explains our observation of the excitonic insulating state only near an intermediate twist angle near $2^\circ$.

\begin{figure}
\includegraphics[width=1\columnwidth]{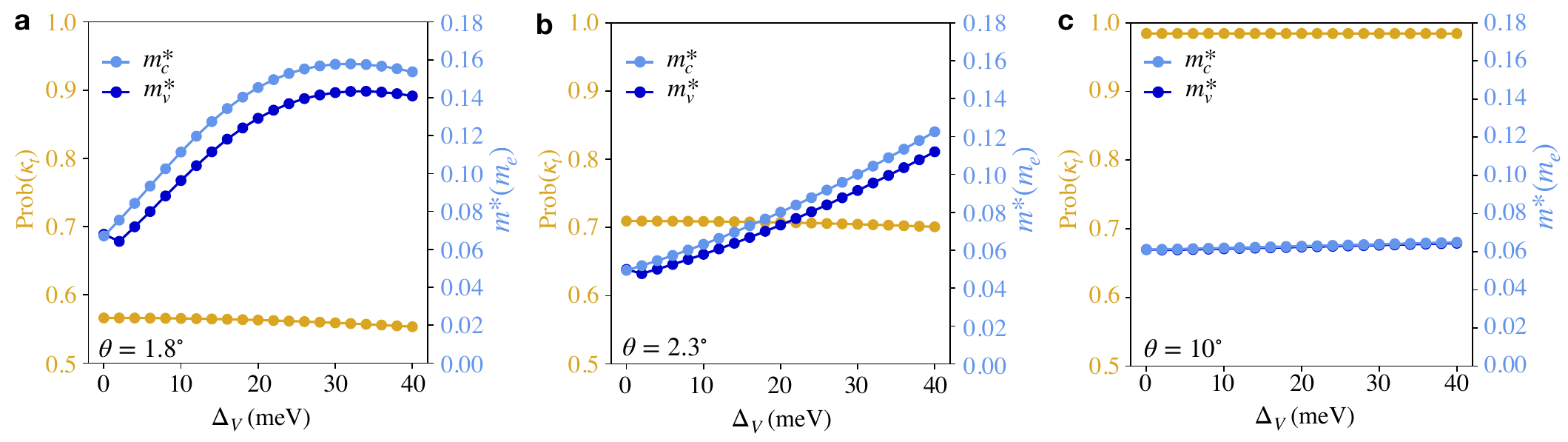}
\centering
\caption{Layer polarization Prob($\kappa_t$) and effective mass $m^*$ as a function of $\Delta_V$ for different twist angles. The crystal fields are absent. (a) $\theta=1.8^\circ$, (b) $\theta=2.3^\circ$, (c) $\theta=10^\circ$.}
\label{suppfig:prob_mass_DeltaV}
\end{figure}

The layer polarization and the effective mass as a function of displacement field parameter $\Delta_\text{V}$ for $\theta=1.8^\circ$, $2.3^\circ$ and $10^\circ$ are shown in Fig.~\ref{suppfig:prob_mass_DeltaV}. As we pointed out in the main text, the layer polarization is almost independent of $\Delta_\text{V}$. For large twist angle, for example $10^\circ$ in Fig.~\ref{suppfig:prob_mass_DeltaV}c, effective masses $m^*_c$ and $m^*_v$ are the same and independent of $\Delta_\text{V}$ as the energy scale of $\Delta_V$ shown in the figure is smaller than the bandwidth. This indicates high symmetry between conduction and valence band. For intermediate twist angles, for example $2.3^\circ$ in Fig.~\ref{suppfig:prob_mass_DeltaV}b, both layer polarization and effective masses are relatively large, which is close to the optimal situation to realize the excitonic insulator. For small twist angles, for example $1.8^\circ$ in Fig.~\ref{suppfig:prob_mass_DeltaV}a, the layer polarization is small even though the effective mass is large. For intermediate twist angles, $m^*_c$ and $m^*_v$ differs by an amount which is increasing with $\Delta_\text{V}$, this partly explains the correlated gap decreasing for large $\Delta_\text{V}$ as in Fig.~\ref{fig:4}b as a result of asymmetric bands.

\subsection{Self-consistent Hartree Fock using an effective two-band model}\label{sec:HF}

In Fig.~\ref{suppfig:probk}b, we have shown that the low-energy bands of $\theta=2.3^\circ$ near Fermi level at charge neutrality mainly localize on the middle two graphene layers on a specific sublattice and a moir\'e reciprocal lattice vector $\mathbf{G}_0 = 0$. We, therefore, can map the Hamiltonian to a $2\times 2$ effective Hamiltonian using the partition technique developed by Feshbach\cite{Feshbach} and Lo\"wdin\cite{Lowdin}. We divide the Hilbert space of TDBG into two parts, target subspace and background subspace with subscripts $t$ and $b$ respectively:
\begin{equation}
H = 
    \begin{pmatrix}
    H_t & H_{tb} \\
    H_{bt} & H_b
    \end{pmatrix}
\end{equation}
After mapping, the effective Hamiltonian in the target subspace $H^{\text{eff}}_t$ keeps the same eigenvalues and eigenvectors as $H$:
\begin{equation}
    H_t^{\text{eff}} = H_t + H_{tb}(\varepsilon_n-H_b)^{-1}H_{bt}
\end{equation}
where $\varepsilon_n$ is eigenvalue of the original Hamiltonian $H$.

The basis of $2 \times 2$ effective Hamiltonian $H^{\text{eff}}$ of TDBG is $(z_{A2,\mathbf{G}_0} \ \  z_{B3,\mathbf{G}_0})^T$, where 2 and 3 are layer indices, $A$ and $B$ are sublattices and moir\'e reciprocal lattice vector $\mathbf{G}_0 = 0$. The Coulomb interaction of this two-band model is
\begin{equation}
    V = \frac{1}{2A} \sum\limits_{\mathbf{k}, \mathbf{k}', \mathbf{q}} \Big(V_q^S a^\dagger_{c,\mathbf{k}+\mathbf{q}} a^\dagger_{c, \mathbf{k}'-\mathbf{q}} a_{c,\mathbf{k}'} a_{c,\mathbf{k}}
    + V_q^S a^\dagger_{v,\mathbf{k}+\mathbf{q}} a^\dagger_{v, \mathbf{k}'-\mathbf{q}} a_{v,\mathbf{k}'} a_{v,\mathbf{k}} \Big) 
    + \frac{1}{A} \sum\limits_{\mathbf{k}, \mathbf{k}', \mathbf{q}} 
    V_q^D a^\dagger_{c,\mathbf{k}+\mathbf{q}} a^\dagger_{v, \mathbf{k}'-\mathbf{q}} a_{v,\mathbf{k}'} a_{c,\mathbf{k}}
\end{equation}
where $c$ and $v$ denote conduction and valence band respectively, A is the sample area. The Coulomb potential in the same layer is $V^S_q = \frac{2\pi e^2}{\epsilon q}$, and the Coulomb potenial between different layers is $V_q^D = \frac{2\pi e^2}{\epsilon q} e^{-qd}$. In the Hartree Fock mean-field theory, the total Hamiltonian is
\begin{eqnarray}
    H_{\text{MF}} = H^{\text{eff}} + V_{\text{MF}} 
    =
    \sum\limits_{\mathbf{k}}
    \begin{pmatrix}
    a^\dagger_{c\mathbf{k}} & a^\dagger_{v\mathbf{k}}
    \end{pmatrix}
    \begin{pmatrix}
    \varepsilon_{c \mathbf{k}} & -\Delta_{\mathbf{k}} \\
    -\Delta^*_{\mathbf{k}} & \varepsilon_{v \mathbf{k}}
    \end{pmatrix}
    \begin{pmatrix}
    a_{c\mathbf{k}} \\ a_{v\mathbf{k}}
    \end{pmatrix}
\end{eqnarray}
A set of equations are solved self consistently:
\begin{eqnarray}
    \label{eq_set1} \varepsilon_{c\mathbf{k}} &=& \varepsilon_{c\mathbf{k}}^0 + \frac{2\pi\varepsilon_{c\mathbf{k}}^0 e^2 dn_e}{\epsilon} - \frac{1}{A} \sum\limits_{\mathbf{k}'} V^S(\mathbf{k}'-\mathbf{k}) \langle a^\dagger_{c\mathbf{k}'} a_{c\mathbf{k}'} \rangle \\
    \label{eq_set2} \varepsilon_{v\mathbf{k}} &=& \varepsilon_{v\mathbf{k}}^0 - \frac{2\pi e^2 dn_h}{\epsilon} - \frac{1}{A} \sum\limits_{\mathbf{k}'} V^S(\mathbf{k}'-\mathbf{k}) \Big(\langle a^\dagger_{v\mathbf{k}'} a_{v\mathbf{k}'} \rangle - 1 \Big) \\
    \label{eq_set3} \varepsilon_{\mathbf{k}} &=& \frac{1}{2} (\varepsilon_{c\mathbf{k}} - \varepsilon_{v\mathbf{k}}) \\
    \label{eq_set4} \Delta_{\mathbf{k}} &=& \frac{1}{A} \sum\limits_{\mathbf{k}'} V^D(\mathbf{k}' - \mathbf{k}) \frac{\Delta_{\mathbf{k}'}}{2\sqrt{\varepsilon^2_{\mathbf{k}'} + \Delta^2_{\mathbf{k}'}}} \\
    \label{eq_set5} n_e &=& \frac{1}{A} \sum\limits_{\mathbf{k}} \langle a^\dagger_{c\mathbf{k}} a_{c\mathbf{k}} \rangle 
\end{eqnarray}
where $\varepsilon_{c\mathbf{k}}^0$ and $\varepsilon_{v\mathbf{k}}^0$ are eigenvalues of $H^{\text{eff}}$.

\subsection{Static polarization function and dielectric constant}\label{sec:dielectricconstant}
The electric field modifies the band structure, especially the band gaps of top and bottom bilayers as shown in Fig.~\ref{suppfig:asymmetry}, thereby the static dielectric constant. We examine the static polarization function using the Lindhard formula
\begin{equation}
    \Pi_0^{\mathbf{G} \mathbf{G}'}(\mathbf{q}, \omega \rightarrow 0) = \frac{g}{A} \sum_{\substack{n,m,\mathbf{k}\\ \mathbf{G}_1, \mathbf{G}_2}}
    \frac{f(\varepsilon_{n,\mathbf{k}}) - f(\varepsilon_{m,\mathbf{k}+\mathbf{q}})}{\varepsilon_{n,\mathbf{k}}- \varepsilon_{m,\mathbf{k}+\mathbf{q}}}  
    \langle \psi_{n,\mathbf{k}+\mathbf{G}_1} | e^{-i(\mathbf{q}+\mathbf{G}) \cdot \mathbf{r}} |\psi_{m,\mathbf{k}+\mathbf{q}+\mathbf{G}_1} \rangle
    \langle \psi_{m,\mathbf{k}+\mathbf{q}+\mathbf{G}_2} | e^{i(\mathbf{q}+\mathbf{G}') \cdot \mathbf{r}} | \psi_{n,\mathbf{k}+\mathbf{G}_2} \rangle
\end{equation}
$\mathbf{G}$, $\mathbf{G}'$, $\mathbf{G}_1$ and $\mathbf{G}_2$ are moir\'e reciprocal lattice vectors, $\mathbf{k}$ and $\mathbf{q}$ are in the first moir\'e Brillouin zone, $A$ is area, $g=4$ includes spin and valley degeneracies.

\begin{figure}
\includegraphics[width=0.6\columnwidth]{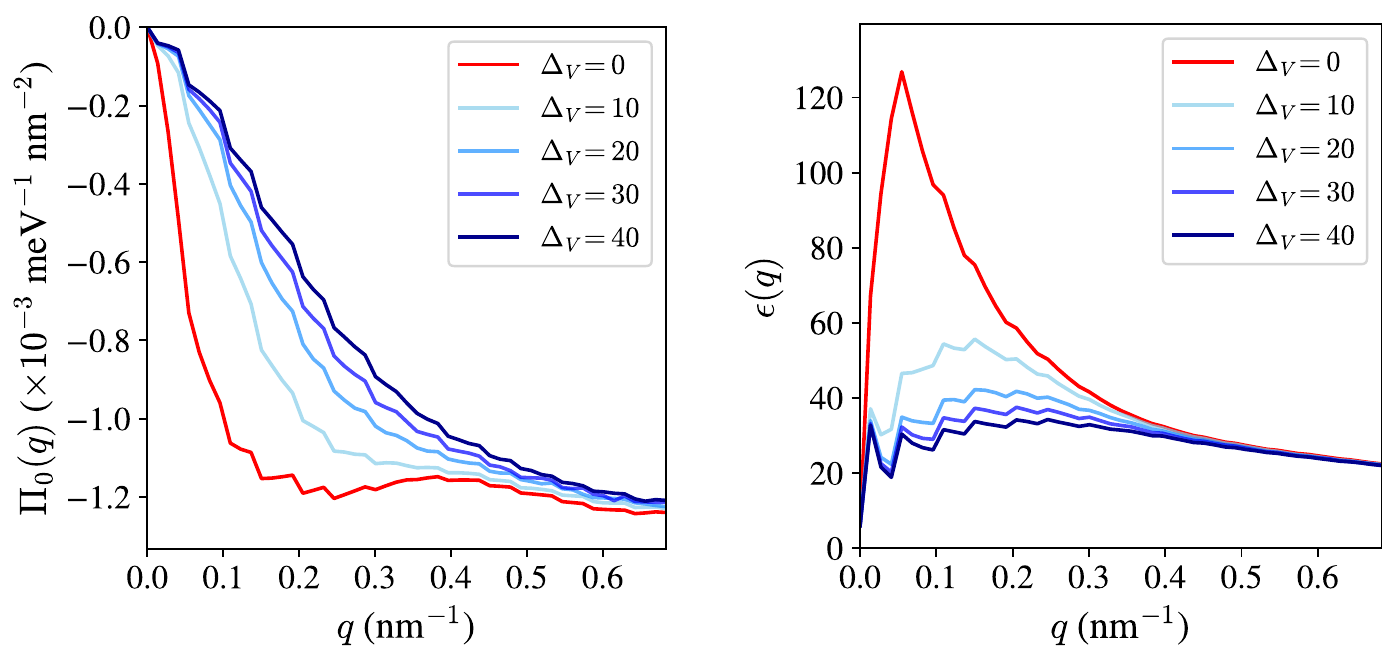}
\centering
\caption{Static polarization function $\Pi_0(q)$ and dielectric function $\epsilon(q)$ of $2.3^\circ$-TDBG for different displacement fields $\Delta_\text{V}=0-40$ meV. The dielectric function decreases with $\Delta_{\text{V}}$.}
\label{suppfig:epsilon}
\end{figure}

The characteristic wavelength of the Fermi surface near charge neutrality is much longer than the moir\'e period, we can therefore consider only the polarization function matrix element of $\mathbf{G} = \mathbf{G}' = 0$:
\begin{equation}
    \Pi_0(\mathbf{q}) = \frac{g}{A} \sum_{\substack{n,m,\mathbf{k}\\ \mathbf{G}_1, \mathbf{G}_2}}
    \frac{f(\varepsilon_{n,\mathbf{k}}) - f(\varepsilon_{m,\mathbf{k}+\mathbf{q}})}{\varepsilon_{n,\mathbf{k}}- \varepsilon_{m,\mathbf{k}+\mathbf{q}}}  
    \langle \psi_{n,\mathbf{k}+\mathbf{G}_1} | e^{-i\mathbf{q} \cdot \mathbf{r}} |\psi_{m,\mathbf{k}+\mathbf{q}+\mathbf{G}_1} \rangle
    \langle \psi_{m,\mathbf{k}+\mathbf{q}+\mathbf{G}_2} | e^{i\mathbf{q} \cdot \mathbf{r}} | \psi_{n,\mathbf{k}+\mathbf{G}_2} \rangle
\end{equation}
and the static dielectric constant stemmed from the interband transitions is
\begin{equation}
    \epsilon_i(\mathbf{q}) = 1-\frac{V(\mathbf{q})}{\epsilon_b} \Pi_0(\mathbf{q})
\end{equation}
$V(\mathbf{q})$ is the bare Coulomb potential and $\epsilon_b$ is the background dielectric constant coming from surrounded dielectrics. In TDBG encapsulated by hBN substrates, we use $\epsilon_b = 6$ for hBN.
The total dielectric constant incorporated in self-consistent Hartree Fock calculation is
\begin{equation}\label{tot_epsilon}
    \epsilon(\mathbf{q}) = \epsilon_b - V(\mathbf{q}) \Pi_0(\mathbf{q}) 
\end{equation}
We show the static polarization function $\Pi_0(q)$ and dielectric constant $\epsilon(q)$ of $2.3^\circ$-TDBG for different $\Delta_\text{V}$ in Fig.~\ref{suppfig:epsilon}.

\end{document}